\documentclass[aps,prl,reprint,twocolumn,superscriptaddress,amsmath,amssymb,citeautoscript]{revtex4-1}
\usepackage{graphicx}
\usepackage{dcolumn}
\usepackage{bm}
\usepackage{latexsym,epsfig}
\usepackage{graphicx}
\usepackage{verbatim}
\usepackage{comment}
\usepackage{amsmath}
\usepackage{amssymb}
\usepackage{color}
\usepackage{epstopdf}
\usepackage{grffile}
\DeclareGraphicsExtensions{.eps}

\newcommand{\beq}{\begin{equation}}
\newcommand{\eeq}{\end{equation}}
\newcommand{\bea}{\begin{eqnarray}}
\newcommand{\eea}{\end{eqnarray}}
\newcommand{\ben}{\begin{eqnarray*}}
\newcommand{\een}{\end{eqnarray*}}
\newcommand{\bfig}{\begin{figure}}
\newcommand{\efig}{\end{figure}}

\newcommand{\ra}{\rangle}
\newcommand{\la}{\langle}

\begin{document}
\title{Topological charge pumping of bound bosonic pairs}
\author{Sebastian Greschner}
\affiliation{Department of Quantum Matter Physics, University of Geneva, 1211 Geneva, Switzerland}
\author{Suman Mondal}
\author{Tapan Mishra}
\affiliation{Department of Physics, Indian Institute of Technology, Guwahati-781039, India}

\date{\today}

\begin{abstract}
Experiments with bosonic atoms in optical superlattices allow for the interesting possibility to study the adiabatic quantized pumping of bosonic atoms in the presence of interactions. 
We investigate this exotic phenomenon for bound bosonic pairs in the paradigmatic Su-Schrieffer-Heeger model where the ground 
state exhibits topological phase transitions due to dimerized hoppings. 
At unit filling we show that there exist crossovers and phase transitions to bond-order phases of paired bosons known as pair-bond-order phase 
as a function of attractive interactions. The pair bond order phase is found to exhibit effective topological properties such as the presence of polarized paired edge states. 
This is further analyzed by studying the emergence and breakdown of the Thouless charge pumping of this bound bosonic pairs by a parametric extension to the famous Rice-Mele model.
Finally we discuss how the pumping of paired bosons or different regimes of breakdown of charge pumping can be probed by state-of-the 
art experiments with repulsively bound bosons.
\end{abstract}

\maketitle

Composite particles often exhibit fundamentally different properties, e.g. charge or exchange statistics, from the ones of their constituents, which may strongly 
influence the properties of many body states of these composite objects. A paradigmatic example of this is the composite fermions picture of fractional 
quantum Hall states~\cite{Jain1989,Halperin1993}.
In particular, these phases have triggered a paramount interest in topological phases of matter influencing the field of condensed matter physics, 
material sciences and quantum computation~\cite{Hasan2010}. Recently a strong effort has been put to realize such interesting physics in ultracold 
quantum gases experiments~\cite{Lohse2015,Takahashi2016pumping,Schweizer2016,Lohse2018}. 
Many topological quantum phases have been studied extensively in the context of non-interacting fermionic systems in their ground state. During the 
recent years a great deal of progress has been made to extend these concepts to interacting fermions~\cite{Manmana2012, Yoshida2014} and bosons~\cite{Grusdt2013}, 
dynamical fields~\cite{Gonzalez2018, Gonzalez2018b, magnifico2019symmetry}, 
as well as finite temperatures, non-equilibrium and mixed states~\cite{bardyn2013topology, bardyn2018probing, heyl2017dynamical}. 
In this paper we explore the non-trivial topological properties of bosonic composite pairs for example the bound bosonic pairs which may be directly accessible in state-of-the-art quantum gas experiments.

\begin{figure}[b]
\begin{center}
\includegraphics[width=0.99\columnwidth]{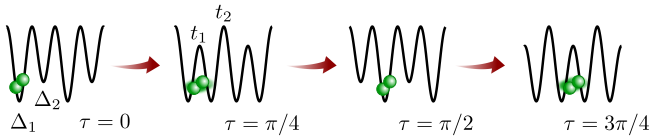}
\end{center}
\caption{Sketch of a charge pumping cycle of the RM model in a one dimensional optical superlattice. 
{The SSH-model corresponds to a dimerized hopping $\delta t = (t_1-t_2)/2$ as shown in the configurations $\tau=\pi/4$ and $\tau=3\pi/4$. 
The cases of $\tau=0$ and $\tau=\pi/2$ depict a staggered potential with $\delta \Delta=\Delta_2-\Delta_1$.} 
}
\label{fig:lattice}
\end{figure}

One of the simplest one dimensional models which possesses non-trivial topological features is the Su-Schrieffer-Heeger~(SSH) model~\cite{ssh} which has been
extensively studied in the context of fermionic and bosonic systems~\cite{Grusdt2013,Ryu2002,Delplace2011,Lang2012,Manmana2012,Yoshida2014,Atala2013,DiLiberto2016,DiLiberto2017,Leseleuc2018}.
These topological phases in the SSH model are characterized by the existence of polarized edge states which may be probed by the presence of adiabatic transport or pumping of a quantized topological charge.
This concept was first introduced by Thouless~\cite{Thouless1983}, and may be studied by the generalization of the SSH to the Rice-Mele~(RM) model~\cite{Rice1982}. 
Recently, with the observation of charge pumping in cold-atom experiments \cite{Lohse2015,Takahashi2016pumping,Schweizer2016,Lohse2018}, the fate of Thouless-pumping in 
interacting systems, such as the interacting fermionic or bosonic RM model has attracted a lot of interest~\cite{Berg2011, Kraus2012pumping,Taddia2017,Nakagawa2018,Hayward2018}. 

In this paper we theoretically investigate the topological phase transitions and Thouless pumping for the bosonic pairs in the context of a generalized RM model given by (compare Fig.~\ref{fig:lattice}) 
\begin{align}
{H}_{\rm RM} =& -\sum_i (t - (-1)^i \delta t \cos(2 \tau)) a_i^\dagger a_{i+1} + {\rm H.c.} \,+\,\nonumber\\
&+ \frac{\delta \Delta}{2} \sin(2 \tau) \sum_i (-1)^i n_i\,+H_{int}.
\label{eq:HRM}
\end{align}
with $a_i^{(\dagger)}$ being the bosonic annihilation(creation) operators on site $i$ and $n_i=a_i^{\dagger}a_i^{\phantom \dagger}$ is the number operator. 
$\tau$ is a cyclic parameter which will be utilized for the pumping protocol. Onsite interactions $H_{\rm int}= \frac{U}{2}\sum_{i}  n_i(n_i-1)$ are characterized by the term $U$.
Note, that the RM model reduces for $\tau=\pi/4$ and $3\pi/4$ to the SSH model,
\begin{align}
{\rm H}_{\rm SSH} = - t \sum_i (1+\delta t (-1)^i) a^\dagger_i a_{i+1} + \text{H.c.} + H_{\rm int}
\label{eq:HSSH}
\end{align}
with staggered hopping rates $t_1$ and $t_2$ from odd and even sites respectively ($t=\frac{t_1+t_2}{2}$ and $\delta t=\frac{t_1-t_2}{2}$). 
We exploit this scenario to first draw insights about the topological phase transitions of 
the bosonic pairs in the SSH model and then analyze the charge pumping of composite pairs in the RM model in one dimension. 
As we want to study the case of bound bosonic pairs we impose three-body 
constraint i.e. $(a_i^{\dagger})^3=0$ in order to stabilize the system against collapse due to the attractive onsite interactions - below we 
discuss how experiments may realize this physics without this constraint.

\begin{figure}[tb]
  \centering
  \includegraphics[height=4.7cm]{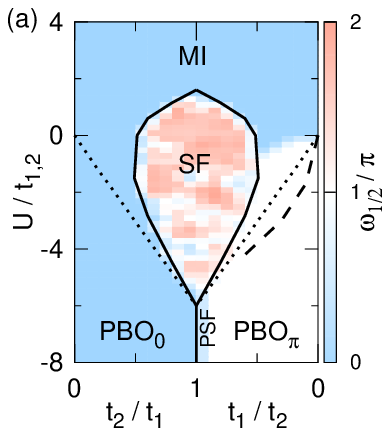}
  \includegraphics[height=4.7cm]{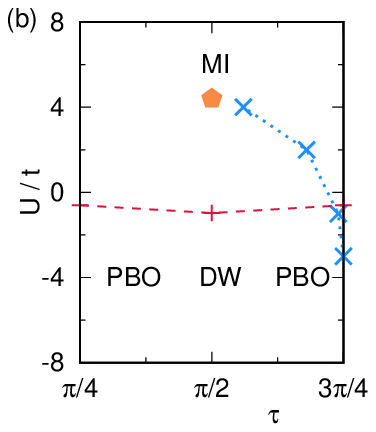}
    \caption{(a) Phase diagram of the bosonic 3-body constraint SSH model at unit filling $n=1$ as function of $t_2/t_1$ and $U/t_1$ (left) and  $t_1/t_2$ and $U/t_2$ (right). 
    DMRG-simulations reveal the BKT phase transitions to the SF phases (solid lines), MI-PBO cross-over (dotted line) and emergence of polarized edge states (dashed line). The gapless 
    PSF phase separates different PBO phases(see Ref.~\cite{supmat}). 
    Colors depict the estimate of the winding numbers $\omega_{1/2}$ (ED, $L=8$).
(b) Charge pumping in Model~\eqref{eq:HRM} for $\delta t=0.9$, $\delta \Delta=2$. The crosses mark the breakdown of 
pair-pumping as seen by the sharp kink in the polarization, red plus symbols the crossing position of two- and single particle excitations. Only for $\tau=\pi/2$ we observe a gapless phase transition 
point between a density wave(DW) and a MI like region (orange hexagon).
}
    \label{fig:pd}
\end{figure}

{\em SSH model --} In the SSH model two types of hopping dimerization are possible for $\tau=\pi/4$ and $3\pi/4$ (Fig.~\ref{fig:lattice}) corresponding to 
$t_1 > t_2$ and $t_1 < t_2$, which exhibit identical bulk properties.
At half filling the single particle spectrum is gapped for any imbalance in hopping between the unit cells $t_1\neq t_2$~\cite{ssh}. 
In the limit of large interactions $U\rightarrow \infty$, the bosons are hardcore in nature and in this limit 
the Model~(\ref{eq:HSSH}), after a Jordan-Wigner transformation to free fermions $c_i^{(\dagger)}$ can be considered as the topological 
SSH model as mentioned before.
Hence, one gets a bond order (BO) phase of bosons at half filling i.e. $n=N/L=1/2$ particles per lattice site as a result of natural dimerization due to the Peierls instability. 
Presence of chiral symmetry in the model leads to the emergence of gapless topological edge states for $t_1<t_2$ which are characterized 
by a nontrivial winding number (or Zak phase)~\cite{Zak1989,Delplace2011,Lang2012}.
This paradigmatic example of bulk-boundary correspondence can be extended to the case of softcore bosons at half filling~\cite{Grusdt2013}. 
Detailed ground state bulk properties of Model~\eqref{eq:HSSH} in a grand canonical ensemble for three-body constrained bosons are discussed in Ref.~\cite{Mondal2019}. 
Here, we will discuss the topological phase transitions as a function of $U$ at unit-filling $n=1$. 

\begin{figure}[tb]
\begin{center}
\includegraphics[width=0.49\linewidth]{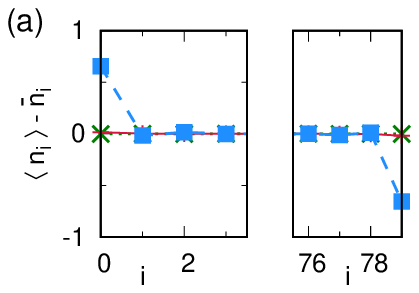}
\includegraphics[width=0.49\linewidth]{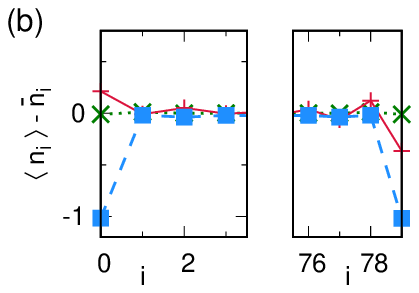}
\end{center}
\caption{Polarized edge states for the RM model at (a) $U=-10$ and (b) $U=-2$ ($t_1=1,~t_2=0.2$) after subtracting the overall 
density modulation $\bar{n}_i = \la n_{L/2 + (i\, {\rm mod}\, 2)} \ra$ for clarity. The different curves are for different 
values of $\tau=0.0$(blue dash-square), $0.25\pi$(red plus-solid), $0.5\pi$(green dot-cross).
}
\label{fig:edges}
\end{figure}

The case of $\delta t=0$ of Model~\eqref{eq:HSSH} is known to exhibit a Berezinskii-Kosterlitz-Thouless(BKT) type phase transition from a gapped Mott-insulator~(MI) to a gapless superfluid phase~(SF) 
for strong repulsive interactions $U>U_c>0$. For negative values of the interaction $U<0$, the 3-body constrained system exhibits an Ising-type 
phase transition to a superfluid phase of paired bosons called the pair superfluid(PSF) phase. 
In the limit of strong attractions $-U \gg t$ the system is a stable ensemble 
of bosonic pairs $(a_i^\dagger)^2|0\rangle$~\cite{Daley2009}. 
Interestingly, this regime is dual to the hardcore boson limit $U\to\infty$ with renormalized 
hopping coefficients of bound pairs $t_i^{eff}=t_i^2/|U|$. For $\delta t>0$ an excitation gap opens up by moving away from the gapless SF and PSF phases as can 
be shown by a field theoretical treatment~\cite{supmat}. 
The complete ground state phase diagram of the three-body constrained SSH model, computed using the density matrix renormalization group~(DMRG) method, 
is shown in Fig.~\ref{fig:pd}~(a) as a function of the hopping ratios $t_2/t_1$ as well as $t_1/t_2$ and the interaction strengths $U/t_1$ and $U/t_2$ respectively. 
For large values of $\delta t$, we observe a smooth crossover from the MI region on the repulsive
$U$ regime to the pair bond order~(PBO) phases in the attractive regime, without closing of any excitation gap.  We may characterize the crossover between MI and 
PBO region by the crossing of 2-particle and 1-particle excitations (dotted line in Fig.~\ref{fig:pd}~(a)~\cite{supmat}).

{\em Edge states --} In our numerical simulations we observe the presence  of polarized edges at finite values of $U/t_2>-\infty$ in the region marked as PBO$_\pi$ phase in Fig.~\ref{fig:pd}~(a). 
In Fig.~\ref{fig:edges} we sketch the edge density for a generalized RM model at $t_1/t_2=0.2$ at different values of $U$~\cite{Backgroundnote}.  
The $\tau=0$ curves show the edge density of the corresponding SSH model. 
For smaller values of $U$ (Fig.~\ref{fig:edges}~(b)) the polarization vanishes abruptly and symmetric (Friedel-like) density oscillations are found at the boundaries of the system.
These edge properties are further quantified by calculating the polarization $P = \frac{1}{L} \sum_{i=0}^{L} \langle \psi | (i-i_0) n_i | \psi \rangle$
with $i_0=(L-1)/2$ and the ground state $|\psi\rangle$ of Model~\eqref{eq:HRM} or ~\eqref{eq:HSSH}~\cite{supmat}.

The concept of bulk-boundary correspondence states that the presence of topological edge states should be related to a 
non-trivial topological invariant of the bulk system. For the SSH-model this is the winding number~\cite{Grusdt2013} defined 
in the  many-body context as $\omega = \int_0^{2\pi} d\theta \langle \psi(\theta) |\partial_\theta \psi(\theta) \rangle$ 
from the ground state $|\psi\rangle$ of the 
effective model with twisted boundary conditions $a_{i} \to {\rm e}^{i\theta / L} a_i$.
With this definition we find that $\omega$ vanishes in all the gapped phases and in the SF phase it is 
accurately characterized by the winding number $\omega\neq 0$ corresponding to its superfluid density. 
However, we observe no distinction between the $t_1<t_2$ and the  $t_2<t_1$ region~\cite{supmat}. 

To circumvent this we may identify $c_i \to (a_i^\dagger)^2$ and, hence, a single fermion hopping 
corresponds to two boson tunneling and will involve twice the phase. 
This simple argument already explains why we observe $\omega=0$ also for $t_2<t_1$ as 
we are winding effectively twice around the parameter space.
Hence, the winding number correctly describing the topological properties in this limit should be defined over half the period as
\begin{align}
\omega_{1/2} = \int_0^{\pi} d\theta \langle \psi(\theta) |\partial_\theta \psi(\theta) \rangle \,,
\label{eq:winding12}
\end{align}
in analogy to a $Z_2$ index used for the description of e.g. quantum spin Hall effect~\cite{kane2005}.
In Fig.~\ref{fig:pd}~(a) we calculate $\omega_{1/2}$ for the full phase diagram for small system sizes and observe that, interestingly, $\omega_{1/2}$ is accurately 
quantized in all gapped regions. We also observe an extended region $\omega_{1/2}=\pi$ which coincides roughly with the emergence of the edge states. This allows us to discriminate between the $PBO_0$ and a $PBO_\pi$ regions with non-trivial and trivial effective topology, $\omega_{1/2}=0$ and $\omega_{1/2}=\pi$ and an abrupt jump between them, even though both regions remain adiabatically connected in the bulk.

{\em Charge pumping --} We will now extend the discussion on the topological properties of the SSH model to the case of the RM model~\eqref{eq:HRM} which 
connects the $t_1<t_2$ and the $t_2<t_1$ region of the SSH model by a periodic process in the cyclic parameter $\tau$ realizing a Thouless charge pump.
In the single particle picture (i.e. for $U\to -\infty$) the pumped charge can be related to a Chern number of the RM model (in momentum $k$ and $\tau$ space). 
Hence, here the pumped charge is quantized and directly linked to the non-trivial topology of the model if non-vanishing. 
\begin{figure}[tb]
  \centering
\begin{minipage}{.48\linewidth}
\includegraphics[width=0.99\linewidth]{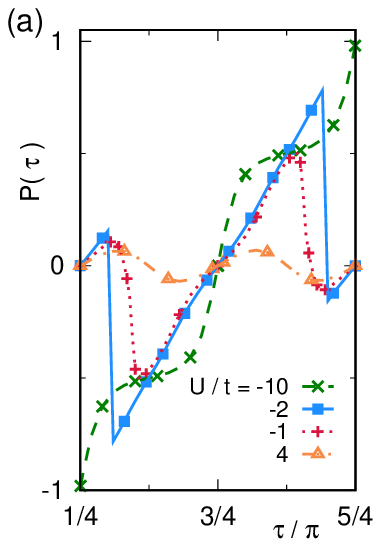}
\end{minipage}
\begin{minipage}{.48\linewidth}
\vspace{-0.1cm}
\includegraphics[width=0.99\linewidth]{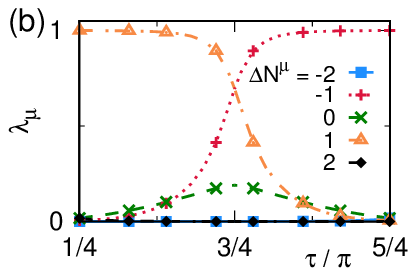}
\includegraphics[width=0.99\linewidth]{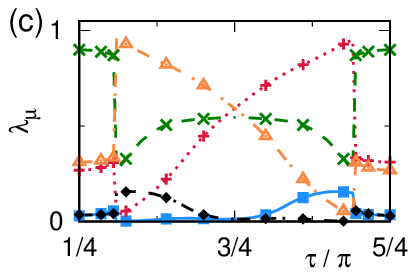}
\end{minipage}
    \caption{(a) Polarization $P$ for the RM model~\eqref{eq:HRM} as function of the adiabatic parameter $\tau$ for 
    several values of $U$ ($\delta \Delta=\delta t$). 
    Note that, $\tau=\pi/4$ and $5\pi/4$ correspond to $t_1=0.2$ and $t_2=1$ and for $\tau=3\pi/4$ we have $t_1=1.0$ and $t_2=0.2$. We consider system 
    size of $L=180$ sites {and for comparison we consider $L=80$(not shown) for $U=-1$ and observe no change in the result}. 
    For $U=4$ we plot $10\times P$ for clarity. (b-c) Largest values of the entanglement spectrum $\lambda_\mu$ of the reduced density matrix in the center of a finite system of size $L=160$ 
   for the model~\eqref{eq:HRM} as a function of the adiabatic parameter $\tau$ for (b) $U=-10$, (c) $U=-2$ (compare Ref.~\cite{Hayward2018}). Note that we show less data points than the calculated ones for 
   better visibility. 
     }
    \label{fig:n1_ppump}
\end{figure}
Following Ref.~\cite{Nakagawa2018} we study the charge pumping for finite systems with open-boundary conditions by monitoring the polarization 
$P(\tau)$ of Model~\eqref{eq:HRM}. The total transferred charge is given by $Q = \int_0^1 d\tau \partial_\tau P(\tau)$
and hence, directly linked to the presence of polarized edge state for the SSH-model. We plot the polarization over the pumping-cycle in Fig.~\ref{fig:n1_ppump}~(a) 
for several values of the interactions.
While for strong attractive interactions $U=-10t$, we observe a clear pumping of a charge $Q=2$ corresponding 
to a bosonic pair, for $U=-2, ~-1$ and $4$ we find zero pumped charge i.e. $Q=0$, corresponding to the abrupt vanishing of the polarized edge states discussed in the previous section. Remarkably, in this case 
we cannot link the breakdown of the charge-pumping to a gap-closing in the pumping cycles.

As discussed recently by Hayward et al.~\cite{Hayward2018} the charge-pumping in the RM model may as well be visualized by the evolution of the entanglement 
spectrum $\lambda_\mu$. In Fig.~\ref{fig:n1_ppump}(b,~c) we plot the largest eigenvalues $\lambda_\mu$ of the reduced density matrix in the center of the system. 
Due to the total particle number conservation of the model the eigenvalues may be labeled by $\Delta N_\mu = N_\mu - N_0$, where $N_\mu$ corresponds to the quantum number of the 
eigenvalues $\lambda_\mu$ and $N_0 = N/2$.
As shown in Fig.~\ref{fig:n1_ppump}~(b) for strong attractive interaction $U=-10t$, the $\Delta N_\mu=\pm 1$ eigenstates dominate, 
leading to a non-zero pumped charge. With increasing $U$ we observe a crossover to the MI regime where the $\Delta N_\mu=0$ state has the largest contribution and no charge is pumped 
(Fig~\ref{fig:n1_ppump}~(c)). 

While so far we have discussed the case $\delta \Delta=\delta t$, Fig.~\ref{fig:pd}~(b) shows the results of a similar analysis of more asymmetric  pumping 
parameters, $\delta t=0.9$ and $\delta \Delta=2.0$, in a phase diagram showing the pumping cycle as function of $U$. We observe pumping of bosonic pairs for strong attractions of $-U \lesssim 3.5t$. 
Interestingly, the transition region of the breakdown of pumping gets extended and can be observed up to $U\sim 4t$.
The blue cross-dotted line depicts the positions where we observe a sharp 
breakdown of charge pumping. For $\tau=\pi/2$ the system exhibits a gapless Gaussian transition point~\cite{tsukano1998berezinskii,tsukano1998spin}. 
Note, that even though the excitation gap becomes small around the Gaussian transition point we do not find any other gapless phases for the given parameters.
Approaching the Gaussian transition point the breakdown of pumping becomes more smooth such that we cannot identify a precise point of pumping breakdown.
As a topological feature charge pumping should be to some extent robust with alterations of the actual pumping protocol or periodic path chosen through the parameter space of the RM model. 
We show this feature by repeating the above analysis for a more elongated path~\cite{supmat}. 

\begin{figure}[tb]
  \centering
  \includegraphics[width=0.49\linewidth]{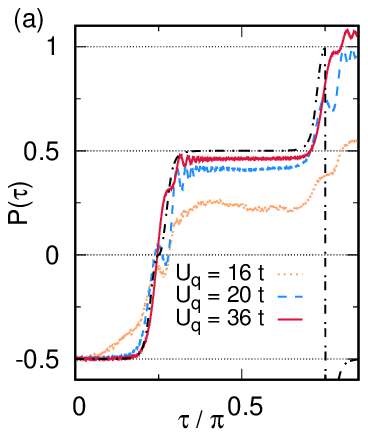}
  \includegraphics[width=0.49\linewidth]{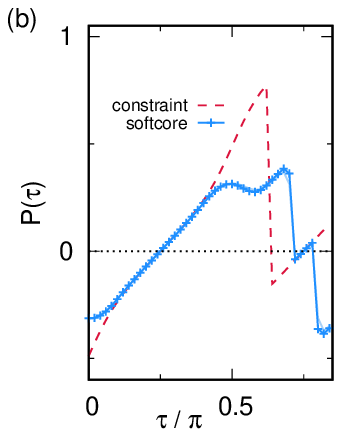}
    \caption{(a) Pumping of repulsively bound pairs for different values of the quenched interactions $U_q=16t$, $20t$ and $36t$ (real-time evolution, $L=6$, $N=6$). 
    The dash-dotted line depicts the fully adiabatic evolution of the effective model of  attractively bound pairs ($L=120$ sites).
(b) Observation of the different regimes of breakdown of charge pumping for bosons with repulsive interactions ($\delta t=0.9$ and $\delta D=2.0$). 
Pumping of three body constrained and softcore bosons ($n_{max}=6$, $L=80$ sites) for $U=3$.}
    \label{fig:tevP_Uneg}
\end{figure}

{\em Experimental realization --} While there has been an active research on the bosonic systems with three body constraint and attractive 
interactions~\cite{baranovprl,petrov1,petrov2,tiesinga,Daley2009,sansone,greschner2015anyon},
important features of the pair-pumping and its breakdown can be studied with state-of-the-art bosonic quantum gas experiments without these properties.
In particular, the pumping of attractive pairs can be simulated by repulsively bound particles: For a deep staggered potential, 
such as shown in Fig.~\ref{fig:pd}~(b) ($\delta t=0.9$ and $\delta \Delta=2.0$ for $\tau=\pi/2$), the ground state  
with good accuracy given by a Fock-state of two bosons in every second lattice site. In a deep optical superlattice this state 
can be accurately prepared~\cite{Bing2019staggered} with unconstrained bosons with a small repulsive interaction $U>0$. After 
initialization, we assume a quench to large repulsive interactions $U_q \gg t,~\delta t,~\delta \Delta$ by means 
a Feshbach resonance. These repulsively bound pairs as studied in Refs.~\cite{winkler2006repulsively,petrosyan2007quantum} 
are stable due to energy conservation and propagate with a reduced hopping rate in a second order tunneling process $t_{eff} \sim 2 t^2/U$. 
One may now try to perform a pumping cycle with these repulsively bound pairs which can simulate the physics of the attractively 
bound bosonic pairs. Note that the pumping process has to be slow compared to the effective tunneling rate but fast enough compared 
to the effective lifetime of the pairs. 
In Fig.~\ref{fig:tevP_Uneg} we study this protocol by means of exact-diagonalization~(ED) simulations and compare to the pumping of attractively bound pairs. 
The effective adiabaticity condition may depend strongly on the precise path through the phase space chosen during the time of evolution~\cite{supmat}. 
For the given examples, $U_q\gtrsim 20 t$ is sufficient to pump one pair during the time-evolution. 
Interestingly, we observe that slightly lower values of $U_q$ quickly lead to a completely distinct evolution without quantized pumped charges.

The choice of more asymmetric pumping parameters, such as the ones of in Fig.~\ref{fig:pd}~(b), shifts a large part of the region of breakdown of pumping 
(blue dot-cross line in Fig.~\ref{fig:pd}~(b)) to positive values of the interaction parameter $U \gtrsim t$.
Here, the three-body constrained system is already to good extent modeled by an unconstrained 
bosonic quantum gas, allowing for the experimental study of the interesting transition region without the 3-body constraint. 
In Fig.~\ref{fig:tevP_Uneg}(b) we compare both cases of constrained and unconstrained bosons for $U=3t$. While both curves differ 
strongly around $\tau=\pi/2$, we observe also for the softcore bosons the sudden kink in the pumped polarization for 
some value $\tau = \tau_c < 3\pi/4$, which we could identify with the breakdown of pair-pumping. 

{\em In summary}, we have investigated the ground-state phase-diagram and topological properties of attractive bosons in the context of 
the SSH and RM model at unit filling. 
For strong attractive interactions the bosons pair up and dimerize to form the PBO phases, with different effective topological properties and winding 
numbers $\omega_{1/2}=0$ and $\omega_{1/2}=\pi$, being linked to interesting edge states of paired bosons.
While aspects of the topological pumping of bosonic pairs could be reproduced with current set-ups of unconstrained bosons, 
the flexibility to tune the interactions from attractive to repulsive regimes and 
the techniques to engineer three and higher order local interactions~\cite{baranovprl,petrov1,petrov2,tiesinga,Daley2009,sansone} and recent experimental observation ~\cite{will2010}
have broadened the scope of simulating the physics of many-body systems by several folds. 
With the existing state of the art facilities the current prediction can in principle be experimentally accessible in ultracold atom experiments along 
the line of recent experiment on Rydberg atoms in SSH model~\cite{Leseleuc2018}.

\begin{acknowledgments}
S.G. acknowledges important discussions with Thierry Giamarchi, Charles-E.~Bardyn, Han-Ning Dai and Zhen-Sheng Yuan as well as financial support by the Swiss National Science Foundation under Division II. 
T.M. acknowledges SERB(India) for the early career grant through Project No.  ECR/2017/001069.
The  computational simulations were carried out using the Param-Ishan HPC facility at Indian Institute of Technology - Guwahati, India and the baobab cluster at University of Geneva.
\end{acknowledgments}

\bibliography{references}

\begin{thebibliography}{60}%
\makeatletter
\providecommand \@ifxundefined [1]{%
 \@ifx{#1\undefined}
}%
\providecommand \@ifnum [1]{%
 \ifnum #1\expandafter \@firstoftwo
 \else \expandafter \@secondoftwo
 \fi
}%
\providecommand \@ifx [1]{%
 \ifx #1\expandafter \@firstoftwo
 \else \expandafter \@secondoftwo
 \fi
}%
\providecommand \natexlab [1]{#1}%
\providecommand \enquote  [1]{``#1''}%
\providecommand \bibnamefont  [1]{#1}%
\providecommand \bibfnamefont [1]{#1}%
\providecommand \citenamefont [1]{#1}%
\providecommand \href@noop [0]{\@secondoftwo}%
\providecommand \href [0]{\begingroup \@sanitize@url \@href}%
\providecommand \@href[1]{\@@startlink{#1}\@@href}%
\providecommand \@@href[1]{\endgroup#1\@@endlink}%
\providecommand \@sanitize@url [0]{\catcode `\\12\catcode `\$12\catcode
  `\&12\catcode `\#12\catcode `\^12\catcode `\_12\catcode `\%12\relax}%
\providecommand \@@startlink[1]{}%
\providecommand \@@endlink[0]{}%
\providecommand \url  [0]{\begingroup\@sanitize@url \@url }%
\providecommand \@url [1]{\endgroup\@href {#1}{\urlprefix }}%
\providecommand \urlprefix  [0]{URL }%
\providecommand \Eprint [0]{\href }%
\providecommand \doibase [0]{http://dx.doi.org/}%
\providecommand \selectlanguage [0]{\@gobble}%
\providecommand \bibinfo  [0]{\@secondoftwo}%
\providecommand \bibfield  [0]{\@secondoftwo}%
\providecommand \translation [1]{[#1]}%
\providecommand \BibitemOpen [0]{}%
\providecommand \bibitemStop [0]{}%
\providecommand \bibitemNoStop [0]{.\EOS\space}%
\providecommand \EOS [0]{\spacefactor3000\relax}%
\providecommand \BibitemShut  [1]{\csname bibitem#1\endcsname}%
\let\auto@bib@innerbib\@empty
\bibitem [{\citenamefont {Jain}(1989)}]{Jain1989}%
  \BibitemOpen
  \bibfield  {author} {\bibinfo {author} {\bibfnamefont {J.~K.}\ \bibnamefont
  {Jain}},\ }\href {\doibase 10.1103/PhysRevLett.63.199} {\bibfield  {journal}
  {\bibinfo  {journal} {Phys. Rev. Lett.}\ }\textbf {\bibinfo {volume} {63}},\
  \bibinfo {pages} {199} (\bibinfo {year} {1989})}\BibitemShut {NoStop}%
\bibitem [{\citenamefont {Halperin}\ \emph {et~al.}(1993)\citenamefont
  {Halperin}, \citenamefont {Lee},\ and\ \citenamefont {Read}}]{Halperin1993}%
  \BibitemOpen
  \bibfield  {author} {\bibinfo {author} {\bibfnamefont {B.~I.}\ \bibnamefont
  {Halperin}}, \bibinfo {author} {\bibfnamefont {P.~A.}\ \bibnamefont {Lee}}, \
  and\ \bibinfo {author} {\bibfnamefont {N.}~\bibnamefont {Read}},\ }\href
  {\doibase 10.1103/PhysRevB.47.7312} {\bibfield  {journal} {\bibinfo
  {journal} {Phys. Rev. B}\ }\textbf {\bibinfo {volume} {47}},\ \bibinfo
  {pages} {7312} (\bibinfo {year} {1993})}\BibitemShut {NoStop}%
\bibitem [{\citenamefont {Hasan}\ and\ \citenamefont {Kane}(2010)}]{Hasan2010}%
  \BibitemOpen
  \bibfield  {author} {\bibinfo {author} {\bibfnamefont {M.~Z.}\ \bibnamefont
  {Hasan}}\ and\ \bibinfo {author} {\bibfnamefont {C.~L.}\ \bibnamefont
  {Kane}},\ }\href {\doibase 10.1103/RevModPhys.82.3045} {\bibfield  {journal}
  {\bibinfo  {journal} {Rev. Mod. Phys.}\ }\textbf {\bibinfo {volume} {82}},\
  \bibinfo {pages} {3045} (\bibinfo {year} {2010})}\BibitemShut {NoStop}%
\bibitem [{\citenamefont {{Lohse M.}}\ \emph {et~al.}(2015)\citenamefont
  {{Lohse M.}}, \citenamefont {{Schweizer C.}}, \citenamefont {{Zilberberg
  O.}}, \citenamefont {{Aidelsburger M.}},\ and\ \citenamefont {{Bloch
  I.}}}]{Lohse2015}%
  \BibitemOpen
  \bibfield  {author} {\bibinfo {author} {\bibnamefont {{Lohse M.}}}, \bibinfo
  {author} {\bibnamefont {{Schweizer C.}}}, \bibinfo {author} {\bibnamefont
  {{Zilberberg O.}}}, \bibinfo {author} {\bibnamefont {{Aidelsburger M.}}}, \
  and\ \bibinfo {author} {\bibnamefont {{Bloch I.}}},\ }\href {\doibase
  http://dx.doi.org/10.1038/nphys3584 10.1038/nphys3584} {\bibfield  {journal}
  {\bibinfo  {journal} {Nature Physics}\ }\textbf {\bibinfo {volume} {12}},\
  \bibinfo {pages} {350} (\bibinfo {year} {2015})}\BibitemShut {NoStop}%
\bibitem [{\citenamefont {Nakajima}\ \emph {et~al.}(2016)\citenamefont
  {Nakajima}, \citenamefont {Tomita}, \citenamefont {Taie}, \citenamefont
  {Ichinose}, \citenamefont {Ozawa}, \citenamefont {Wang}, \citenamefont
  {Troyer},\ and\ \citenamefont {Takahashi}}]{Takahashi2016pumping}%
  \BibitemOpen
  \bibfield  {author} {\bibinfo {author} {\bibfnamefont {S.}~\bibnamefont
  {Nakajima}}, \bibinfo {author} {\bibfnamefont {T.}~\bibnamefont {Tomita}},
  \bibinfo {author} {\bibfnamefont {S.}~\bibnamefont {Taie}}, \bibinfo {author}
  {\bibfnamefont {T.}~\bibnamefont {Ichinose}}, \bibinfo {author}
  {\bibfnamefont {H.}~\bibnamefont {Ozawa}}, \bibinfo {author} {\bibfnamefont
  {L.}~\bibnamefont {Wang}}, \bibinfo {author} {\bibfnamefont {M.}~\bibnamefont
  {Troyer}}, \ and\ \bibinfo {author} {\bibfnamefont {Y.}~\bibnamefont
  {Takahashi}},\ }\href {\doibase http://dx.doi.org/10.1038/nphys3622
  10.1038/nphys3622} {\bibfield  {journal} {\bibinfo  {journal} {Nature
  Physics}\ }\textbf {\bibinfo {volume} {12}},\ \bibinfo {pages} {296}
  (\bibinfo {year} {2016})}\BibitemShut {NoStop}%
\bibitem [{\citenamefont {Schweizer}\ \emph {et~al.}(2016)\citenamefont
  {Schweizer}, \citenamefont {Lohse}, \citenamefont {Citro},\ and\
  \citenamefont {Bloch}}]{Schweizer2016}%
  \BibitemOpen
  \bibfield  {author} {\bibinfo {author} {\bibfnamefont {C.}~\bibnamefont
  {Schweizer}}, \bibinfo {author} {\bibfnamefont {M.}~\bibnamefont {Lohse}},
  \bibinfo {author} {\bibfnamefont {R.}~\bibnamefont {Citro}}, \ and\ \bibinfo
  {author} {\bibfnamefont {I.}~\bibnamefont {Bloch}},\ }\href {\doibase
  10.1103/PhysRevLett.117.170405} {\bibfield  {journal} {\bibinfo  {journal}
  {Phys. Rev. Lett.}\ }\textbf {\bibinfo {volume} {117}},\ \bibinfo {pages}
  {170405} (\bibinfo {year} {2016})}\BibitemShut {NoStop}%
\bibitem [{\citenamefont {Lohse}\ \emph {et~al.}(2018)\citenamefont {Lohse},
  \citenamefont {Schweizer}, \citenamefont {Price}, \citenamefont
  {Zilberberg},\ and\ \citenamefont {Bloch}}]{Lohse2018}%
  \BibitemOpen
  \bibfield  {author} {\bibinfo {author} {\bibfnamefont {M.}~\bibnamefont
  {Lohse}}, \bibinfo {author} {\bibfnamefont {C.}~\bibnamefont {Schweizer}},
  \bibinfo {author} {\bibfnamefont {H.~M.}\ \bibnamefont {Price}}, \bibinfo
  {author} {\bibfnamefont {O.}~\bibnamefont {Zilberberg}}, \ and\ \bibinfo
  {author} {\bibfnamefont {I.}~\bibnamefont {Bloch}},\ }\href@noop {}
  {\bibfield  {journal} {\bibinfo  {journal} {Nature}\ }\textbf {\bibinfo
  {volume} {553}},\ \bibinfo {pages} {55} (\bibinfo {year} {2018})}\BibitemShut
  {NoStop}%
\bibitem [{\citenamefont {Manmana}\ \emph {et~al.}(2012)\citenamefont
  {Manmana}, \citenamefont {Essin}, \citenamefont {Noack},\ and\ \citenamefont
  {Gurarie}}]{Manmana2012}%
  \BibitemOpen
  \bibfield  {author} {\bibinfo {author} {\bibfnamefont {S.~R.}\ \bibnamefont
  {Manmana}}, \bibinfo {author} {\bibfnamefont {A.~M.}\ \bibnamefont {Essin}},
  \bibinfo {author} {\bibfnamefont {R.~M.}\ \bibnamefont {Noack}}, \ and\
  \bibinfo {author} {\bibfnamefont {V.}~\bibnamefont {Gurarie}},\ }\href
  {\doibase 10.1103/PhysRevB.86.205119} {\bibfield  {journal} {\bibinfo
  {journal} {Phys. Rev. B}\ }\textbf {\bibinfo {volume} {86}},\ \bibinfo
  {pages} {205119} (\bibinfo {year} {2012})}\BibitemShut {NoStop}%
\bibitem [{\citenamefont {Yoshida}\ \emph {et~al.}(2014)\citenamefont
  {Yoshida}, \citenamefont {Peters}, \citenamefont {Fujimoto},\ and\
  \citenamefont {Kawakami}}]{Yoshida2014}%
  \BibitemOpen
  \bibfield  {author} {\bibinfo {author} {\bibfnamefont {T.}~\bibnamefont
  {Yoshida}}, \bibinfo {author} {\bibfnamefont {R.}~\bibnamefont {Peters}},
  \bibinfo {author} {\bibfnamefont {S.}~\bibnamefont {Fujimoto}}, \ and\
  \bibinfo {author} {\bibfnamefont {N.}~\bibnamefont {Kawakami}},\ }\href
  {\doibase 10.1103/PhysRevLett.112.196404} {\bibfield  {journal} {\bibinfo
  {journal} {Phys. Rev. Lett.}\ }\textbf {\bibinfo {volume} {112}},\ \bibinfo
  {pages} {196404} (\bibinfo {year} {2014})}\BibitemShut {NoStop}%
\bibitem [{\citenamefont {Grusdt}\ \emph {et~al.}(2013)\citenamefont {Grusdt},
  \citenamefont {H\"oning},\ and\ \citenamefont {Fleischhauer}}]{Grusdt2013}%
  \BibitemOpen
  \bibfield  {author} {\bibinfo {author} {\bibfnamefont {F.}~\bibnamefont
  {Grusdt}}, \bibinfo {author} {\bibfnamefont {M.}~\bibnamefont {H\"oning}}, \
  and\ \bibinfo {author} {\bibfnamefont {M.}~\bibnamefont {Fleischhauer}},\
  }\href {\doibase 10.1103/PhysRevLett.110.260405} {\bibfield  {journal}
  {\bibinfo  {journal} {Phys. Rev. Lett.}\ }\textbf {\bibinfo {volume} {110}},\
  \bibinfo {pages} {260405} (\bibinfo {year} {2013})}\BibitemShut {NoStop}%
\bibitem [{\citenamefont {Gonz\'alez-Cuadra}\ \emph
  {et~al.}(2018{\natexlab{a}})\citenamefont {Gonz\'alez-Cuadra}, \citenamefont
  {Grzybowski}, \citenamefont {Dauphin},\ and\ \citenamefont
  {Lewenstein}}]{Gonzalez2018}%
  \BibitemOpen
  \bibfield  {author} {\bibinfo {author} {\bibfnamefont {D.}~\bibnamefont
  {Gonz\'alez-Cuadra}}, \bibinfo {author} {\bibfnamefont {P.~R.}\ \bibnamefont
  {Grzybowski}}, \bibinfo {author} {\bibfnamefont {A.}~\bibnamefont {Dauphin}},
  \ and\ \bibinfo {author} {\bibfnamefont {M.}~\bibnamefont {Lewenstein}},\
  }\href {\doibase 10.1103/PhysRevLett.121.090402} {\bibfield  {journal}
  {\bibinfo  {journal} {Phys. Rev. Lett.}\ }\textbf {\bibinfo {volume} {121}},\
  \bibinfo {pages} {090402} (\bibinfo {year} {2018}{\natexlab{a}})}\BibitemShut
  {NoStop}%
\bibitem [{\citenamefont {Gonz\'alez-Cuadra}\ \emph
  {et~al.}(2018{\natexlab{b}})\citenamefont {Gonz\'alez-Cuadra}, \citenamefont
  {Dauphin}, \citenamefont {Grzybowski}, \citenamefont {W\'ojcik},
  \citenamefont {Lewenstein},\ and\ \citenamefont {Bermudez}}]{Gonzalez2018b}%
  \BibitemOpen
  \bibfield  {author} {\bibinfo {author} {\bibfnamefont {D.}~\bibnamefont
  {Gonz\'alez-Cuadra}}, \bibinfo {author} {\bibfnamefont {A.}~\bibnamefont
  {Dauphin}}, \bibinfo {author} {\bibfnamefont {P.~R.}\ \bibnamefont
  {Grzybowski}}, \bibinfo {author} {\bibfnamefont {P.}~\bibnamefont
  {W\'ojcik}}, \bibinfo {author} {\bibfnamefont {M.}~\bibnamefont
  {Lewenstein}}, \ and\ \bibinfo {author} {\bibfnamefont {A.}~\bibnamefont
  {Bermudez}},\ }\href@noop {} {\bibfield  {journal} {\bibinfo  {journal}
  {arXiv preprint arXiv:1811.08392}\ } (\bibinfo {year}
  {2018}{\natexlab{b}})}\BibitemShut {NoStop}%
\bibitem [{\citenamefont {Magnifico}\ \emph {et~al.}(2019)\citenamefont
  {Magnifico}, \citenamefont {Vodola}, \citenamefont {Ercolessi}, \citenamefont
  {Kumar}, \citenamefont {M{\"u}ller},\ and\ \citenamefont
  {Bermudez}}]{magnifico2019symmetry}%
  \BibitemOpen
  \bibfield  {author} {\bibinfo {author} {\bibfnamefont {G.}~\bibnamefont
  {Magnifico}}, \bibinfo {author} {\bibfnamefont {D.}~\bibnamefont {Vodola}},
  \bibinfo {author} {\bibfnamefont {E.}~\bibnamefont {Ercolessi}}, \bibinfo
  {author} {\bibfnamefont {S.}~\bibnamefont {Kumar}}, \bibinfo {author}
  {\bibfnamefont {M.}~\bibnamefont {M{\"u}ller}}, \ and\ \bibinfo {author}
  {\bibfnamefont {A.}~\bibnamefont {Bermudez}},\ }\href@noop {} {\bibfield
  {journal} {\bibinfo  {journal} {Physical Review D}\ }\textbf {\bibinfo
  {volume} {99}},\ \bibinfo {pages} {014503} (\bibinfo {year}
  {2019})}\BibitemShut {NoStop}%
\bibitem [{\citenamefont {Bardyn}\ \emph {et~al.}(2013)\citenamefont {Bardyn},
  \citenamefont {Baranov}, \citenamefont {Kraus}, \citenamefont {Rico},
  \citenamefont {{\.I}mamo{\u{g}}lu}, \citenamefont {Zoller},\ and\
  \citenamefont {Diehl}}]{bardyn2013topology}%
  \BibitemOpen
  \bibfield  {author} {\bibinfo {author} {\bibfnamefont {C.}~\bibnamefont
  {Bardyn}}, \bibinfo {author} {\bibfnamefont {M.}~\bibnamefont {Baranov}},
  \bibinfo {author} {\bibfnamefont {C.}~\bibnamefont {Kraus}}, \bibinfo
  {author} {\bibfnamefont {E.}~\bibnamefont {Rico}}, \bibinfo {author}
  {\bibfnamefont {A.}~\bibnamefont {{\.I}mamo{\u{g}}lu}}, \bibinfo {author}
  {\bibfnamefont {P.}~\bibnamefont {Zoller}}, \ and\ \bibinfo {author}
  {\bibfnamefont {S.}~\bibnamefont {Diehl}},\ }\href@noop {} {\bibfield
  {journal} {\bibinfo  {journal} {New Journal of Physics}\ }\textbf {\bibinfo
  {volume} {15}},\ \bibinfo {pages} {085001} (\bibinfo {year}
  {2013})}\BibitemShut {NoStop}%
\bibitem [{\citenamefont {Bardyn}\ \emph {et~al.}(2018)\citenamefont {Bardyn},
  \citenamefont {Wawer}, \citenamefont {Altland}, \citenamefont
  {Fleischhauer},\ and\ \citenamefont {Diehl}}]{bardyn2018probing}%
  \BibitemOpen
  \bibfield  {author} {\bibinfo {author} {\bibfnamefont {C.-E.}\ \bibnamefont
  {Bardyn}}, \bibinfo {author} {\bibfnamefont {L.}~\bibnamefont {Wawer}},
  \bibinfo {author} {\bibfnamefont {A.}~\bibnamefont {Altland}}, \bibinfo
  {author} {\bibfnamefont {M.}~\bibnamefont {Fleischhauer}}, \ and\ \bibinfo
  {author} {\bibfnamefont {S.}~\bibnamefont {Diehl}},\ }\href@noop {}
  {\bibfield  {journal} {\bibinfo  {journal} {Physical Review X}\ }\textbf
  {\bibinfo {volume} {8}},\ \bibinfo {pages} {011035} (\bibinfo {year}
  {2018})}\BibitemShut {NoStop}%
\bibitem [{\citenamefont {Heyl}\ and\ \citenamefont
  {Budich}(2017)}]{heyl2017dynamical}%
  \BibitemOpen
  \bibfield  {author} {\bibinfo {author} {\bibfnamefont {M.}~\bibnamefont
  {Heyl}}\ and\ \bibinfo {author} {\bibfnamefont {J.}~\bibnamefont {Budich}},\
  }\href@noop {} {\bibfield  {journal} {\bibinfo  {journal} {Physical Review
  B}\ }\textbf {\bibinfo {volume} {96}},\ \bibinfo {pages} {180304} (\bibinfo
  {year} {2017})}\BibitemShut {NoStop}%
\bibitem [{\citenamefont {Su}\ \emph {et~al.}(1979)\citenamefont {Su},
  \citenamefont {Schrieffer},\ and\ \citenamefont {Heeger}}]{ssh}%
  \BibitemOpen
  \bibfield  {author} {\bibinfo {author} {\bibfnamefont {W.~P.}\ \bibnamefont
  {Su}}, \bibinfo {author} {\bibfnamefont {J.~R.}\ \bibnamefont {Schrieffer}},
  \ and\ \bibinfo {author} {\bibfnamefont {A.~J.}\ \bibnamefont {Heeger}},\
  }\href {\doibase 10.1103/PhysRevLett.42.1698} {\bibfield  {journal} {\bibinfo
   {journal} {Phys. Rev. Lett.}\ }\textbf {\bibinfo {volume} {42}},\ \bibinfo
  {pages} {1698} (\bibinfo {year} {1979})}\BibitemShut {NoStop}%
\bibitem [{\citenamefont {Ryu}\ and\ \citenamefont {Hatsugai}(2002)}]{Ryu2002}%
  \BibitemOpen
  \bibfield  {author} {\bibinfo {author} {\bibfnamefont {S.}~\bibnamefont
  {Ryu}}\ and\ \bibinfo {author} {\bibfnamefont {Y.}~\bibnamefont {Hatsugai}},\
  }\href {\doibase 10.1103/PhysRevLett.89.077002} {\bibfield  {journal}
  {\bibinfo  {journal} {Phys. Rev. Lett.}\ }\textbf {\bibinfo {volume} {89}},\
  \bibinfo {pages} {077002} (\bibinfo {year} {2002})}\BibitemShut {NoStop}%
\bibitem [{\citenamefont {Delplace}\ \emph {et~al.}(2011)\citenamefont
  {Delplace}, \citenamefont {Ullmo},\ and\ \citenamefont
  {Montambaux}}]{Delplace2011}%
  \BibitemOpen
  \bibfield  {author} {\bibinfo {author} {\bibfnamefont {P.}~\bibnamefont
  {Delplace}}, \bibinfo {author} {\bibfnamefont {D.}~\bibnamefont {Ullmo}}, \
  and\ \bibinfo {author} {\bibfnamefont {G.}~\bibnamefont {Montambaux}},\
  }\href@noop {} {\bibfield  {journal} {\bibinfo  {journal} {Phys. Rev. B}\
  }\textbf {\bibinfo {volume} {84}},\ \bibinfo {pages} {195452} (\bibinfo
  {year} {2011})}\BibitemShut {NoStop}%
\bibitem [{\citenamefont {Lang}\ \emph {et~al.}(2012)\citenamefont {Lang},
  \citenamefont {Cai},\ and\ \citenamefont {Chen}}]{Lang2012}%
  \BibitemOpen
  \bibfield  {author} {\bibinfo {author} {\bibfnamefont {L.-J.}\ \bibnamefont
  {Lang}}, \bibinfo {author} {\bibfnamefont {X.}~\bibnamefont {Cai}}, \ and\
  \bibinfo {author} {\bibfnamefont {S.}~\bibnamefont {Chen}},\ }\href {\doibase
  10.1103/PhysRevLett.108.220401} {\bibfield  {journal} {\bibinfo  {journal}
  {Phys. Rev. Lett.}\ }\textbf {\bibinfo {volume} {108}},\ \bibinfo {pages}
  {220401} (\bibinfo {year} {2012})}\BibitemShut {NoStop}%
\bibitem [{\citenamefont {Atala}\ \emph {et~al.}(2013)\citenamefont {Atala},
  \citenamefont {Aidelsburger}, \citenamefont {Barreiro}, \citenamefont
  {Abanin}, \citenamefont {Kitagawa}, \citenamefont {Demler},\ and\
  \citenamefont {Bloch}}]{Atala2013}%
  \BibitemOpen
  \bibfield  {author} {\bibinfo {author} {\bibfnamefont {M.}~\bibnamefont
  {Atala}}, \bibinfo {author} {\bibfnamefont {M.}~\bibnamefont {Aidelsburger}},
  \bibinfo {author} {\bibfnamefont {J.~T.}\ \bibnamefont {Barreiro}}, \bibinfo
  {author} {\bibfnamefont {D.}~\bibnamefont {Abanin}}, \bibinfo {author}
  {\bibfnamefont {T.}~\bibnamefont {Kitagawa}}, \bibinfo {author}
  {\bibfnamefont {E.}~\bibnamefont {Demler}}, \ and\ \bibinfo {author}
  {\bibfnamefont {I.}~\bibnamefont {Bloch}},\ }\href {\doibase
  https://doi.org/10.1038/nphys2790} {\bibfield  {journal} {\bibinfo  {journal}
  {Nature Physics}\ }\textbf {\bibinfo {volume} {9}},\ \bibinfo {pages} {795}
  (\bibinfo {year} {2013})}\BibitemShut {NoStop}%
\bibitem [{\citenamefont {Di~Liberto}\ \emph {et~al.}(2016)\citenamefont
  {Di~Liberto}, \citenamefont {Recati}, \citenamefont {Carusotto},\ and\
  \citenamefont {Menotti}}]{DiLiberto2016}%
  \BibitemOpen
  \bibfield  {author} {\bibinfo {author} {\bibfnamefont {M.}~\bibnamefont
  {Di~Liberto}}, \bibinfo {author} {\bibfnamefont {A.}~\bibnamefont {Recati}},
  \bibinfo {author} {\bibfnamefont {I.}~\bibnamefont {Carusotto}}, \ and\
  \bibinfo {author} {\bibfnamefont {C.}~\bibnamefont {Menotti}},\ }\href
  {\doibase 10.1103/PhysRevA.94.062704} {\bibfield  {journal} {\bibinfo
  {journal} {Phys. Rev. A}\ }\textbf {\bibinfo {volume} {94}},\ \bibinfo
  {pages} {062704} (\bibinfo {year} {2016})}\BibitemShut {NoStop}%
\bibitem [{\citenamefont {Di~Liberto}\ \emph {et~al.}(2017)\citenamefont
  {Di~Liberto}, \citenamefont {Recati}, \citenamefont {Carusotto},\ and\
  \citenamefont {Menotti}}]{DiLiberto2017}%
  \BibitemOpen
  \bibfield  {author} {\bibinfo {author} {\bibfnamefont {M.}~\bibnamefont
  {Di~Liberto}}, \bibinfo {author} {\bibfnamefont {A.}~\bibnamefont {Recati}},
  \bibinfo {author} {\bibfnamefont {I.}~\bibnamefont {Carusotto}}, \ and\
  \bibinfo {author} {\bibfnamefont {C.}~\bibnamefont {Menotti}},\ }\href
  {\doibase 10.1140/epjst/e2016-60388-y} {\bibfield  {journal} {\bibinfo
  {journal} {The European Physical Journal Special Topics}\ }\textbf {\bibinfo
  {volume} {226}},\ \bibinfo {pages} {2751} (\bibinfo {year}
  {2017})}\BibitemShut {NoStop}%
\bibitem [{\citenamefont {de~L{\'e}s{\'e}leuc}\ \emph
  {et~al.}(2019)\citenamefont {de~L{\'e}s{\'e}leuc}, \citenamefont {Lienhard},
  \citenamefont {Scholl}, \citenamefont {Barredo}, \citenamefont {Weber},
  \citenamefont {Lang}, \citenamefont {B{\"u}chler}, \citenamefont {Lahaye},\
  and\ \citenamefont {Browaeys}}]{Leseleuc2018}%
  \BibitemOpen
  \bibfield  {author} {\bibinfo {author} {\bibfnamefont {S.}~\bibnamefont
  {de~L{\'e}s{\'e}leuc}}, \bibinfo {author} {\bibfnamefont {V.}~\bibnamefont
  {Lienhard}}, \bibinfo {author} {\bibfnamefont {P.}~\bibnamefont {Scholl}},
  \bibinfo {author} {\bibfnamefont {D.}~\bibnamefont {Barredo}}, \bibinfo
  {author} {\bibfnamefont {S.}~\bibnamefont {Weber}}, \bibinfo {author}
  {\bibfnamefont {N.}~\bibnamefont {Lang}}, \bibinfo {author} {\bibfnamefont
  {H.~P.}\ \bibnamefont {B{\"u}chler}}, \bibinfo {author} {\bibfnamefont
  {T.}~\bibnamefont {Lahaye}}, \ and\ \bibinfo {author} {\bibfnamefont
  {A.}~\bibnamefont {Browaeys}},\ }\href {\doibase 10.1126/science.aav9105}
  {\bibfield  {journal} {\bibinfo  {journal} {Science}\ }\textbf {\bibinfo
  {volume} {365}},\ \bibinfo {pages} {775} (\bibinfo {year}
  {2019})}\BibitemShut {NoStop}%
\bibitem [{\citenamefont {Thouless}(1983)}]{Thouless1983}%
  \BibitemOpen
  \bibfield  {author} {\bibinfo {author} {\bibfnamefont {D.}~\bibnamefont
  {Thouless}},\ }\href@noop {} {\bibfield  {journal} {\bibinfo  {journal}
  {Phys. Rev. B}\ }\textbf {\bibinfo {volume} {27}},\ \bibinfo {pages} {6083}
  (\bibinfo {year} {1983})}\BibitemShut {NoStop}%
\bibitem [{\citenamefont {Rice}\ and\ \citenamefont {Mele}(1982)}]{Rice1982}%
  \BibitemOpen
  \bibfield  {author} {\bibinfo {author} {\bibfnamefont {M.}~\bibnamefont
  {Rice}}\ and\ \bibinfo {author} {\bibfnamefont {E.}~\bibnamefont {Mele}},\
  }\href@noop {} {\bibfield  {journal} {\bibinfo  {journal} {Phys. Rev. Lett.}\
  }\textbf {\bibinfo {volume} {49}},\ \bibinfo {pages} {1455} (\bibinfo {year}
  {1982})}\BibitemShut {NoStop}%
\bibitem [{\citenamefont {Berg}\ \emph {et~al.}(2011)\citenamefont {Berg},
  \citenamefont {Levin},\ and\ \citenamefont {Altman}}]{Berg2011}%
  \BibitemOpen
  \bibfield  {author} {\bibinfo {author} {\bibfnamefont {E.}~\bibnamefont
  {Berg}}, \bibinfo {author} {\bibfnamefont {M.}~\bibnamefont {Levin}}, \ and\
  \bibinfo {author} {\bibfnamefont {E.}~\bibnamefont {Altman}},\ }\href@noop {}
  {\bibfield  {journal} {\bibinfo  {journal} {Phys. Rev. Lett.}\ }\textbf
  {\bibinfo {volume} {106}},\ \bibinfo {pages} {110405} (\bibinfo {year}
  {2011})}\BibitemShut {NoStop}%
\bibitem [{\citenamefont {Kraus}\ \emph {et~al.}(2012)\citenamefont {Kraus},
  \citenamefont {Lahini}, \citenamefont {Ringel}, \citenamefont {Verbin},\ and\
  \citenamefont {Zilberberg}}]{Kraus2012pumping}%
  \BibitemOpen
  \bibfield  {author} {\bibinfo {author} {\bibfnamefont {Y.~E.}\ \bibnamefont
  {Kraus}}, \bibinfo {author} {\bibfnamefont {Y.}~\bibnamefont {Lahini}},
  \bibinfo {author} {\bibfnamefont {Z.}~\bibnamefont {Ringel}}, \bibinfo
  {author} {\bibfnamefont {M.}~\bibnamefont {Verbin}}, \ and\ \bibinfo {author}
  {\bibfnamefont {O.}~\bibnamefont {Zilberberg}},\ }\href {\doibase
  10.1103/PhysRevLett.109.106402} {\bibfield  {journal} {\bibinfo  {journal}
  {Phys. Rev. Lett.}\ }\textbf {\bibinfo {volume} {109}},\ \bibinfo {pages}
  {106402} (\bibinfo {year} {2012})}\BibitemShut {NoStop}%
\bibitem [{\citenamefont {Taddia}\ \emph {et~al.}(2017)\citenamefont {Taddia},
  \citenamefont {Cornfeld}, \citenamefont {Rossini}, \citenamefont {Mazza},
  \citenamefont {Sela},\ and\ \citenamefont {Fazio}}]{Taddia2017}%
  \BibitemOpen
  \bibfield  {author} {\bibinfo {author} {\bibfnamefont {L.}~\bibnamefont
  {Taddia}}, \bibinfo {author} {\bibfnamefont {E.}~\bibnamefont {Cornfeld}},
  \bibinfo {author} {\bibfnamefont {D.}~\bibnamefont {Rossini}}, \bibinfo
  {author} {\bibfnamefont {L.}~\bibnamefont {Mazza}}, \bibinfo {author}
  {\bibfnamefont {E.}~\bibnamefont {Sela}}, \ and\ \bibinfo {author}
  {\bibfnamefont {R.}~\bibnamefont {Fazio}},\ }\href {\doibase
  10.1103/PhysRevLett.118.230402} {\bibfield  {journal} {\bibinfo  {journal}
  {Phys. Rev. Lett.}\ }\textbf {\bibinfo {volume} {118}},\ \bibinfo {pages}
  {230402} (\bibinfo {year} {2017})}\BibitemShut {NoStop}%
\bibitem [{\citenamefont {Nakagawa}\ \emph {et~al.}(2018)\citenamefont
  {Nakagawa}, \citenamefont {Yoshida}, \citenamefont {Peters},\ and\
  \citenamefont {Kawakami}}]{Nakagawa2018}%
  \BibitemOpen
  \bibfield  {author} {\bibinfo {author} {\bibfnamefont {M.}~\bibnamefont
  {Nakagawa}}, \bibinfo {author} {\bibfnamefont {T.}~\bibnamefont {Yoshida}},
  \bibinfo {author} {\bibfnamefont {R.}~\bibnamefont {Peters}}, \ and\ \bibinfo
  {author} {\bibfnamefont {N.}~\bibnamefont {Kawakami}},\ }\href {\doibase
  10.1103/PhysRevB.98.115147} {\bibfield  {journal} {\bibinfo  {journal} {Phys.
  Rev. B}\ }\textbf {\bibinfo {volume} {98}},\ \bibinfo {pages} {115147}
  (\bibinfo {year} {2018})}\BibitemShut {NoStop}%
\bibitem [{\citenamefont {Hayward}\ \emph {et~al.}(2018)\citenamefont
  {Hayward}, \citenamefont {Schweizer}, \citenamefont {Lohse}, \citenamefont
  {Aidelsburger},\ and\ \citenamefont {Heidrich-Meisner}}]{Hayward2018}%
  \BibitemOpen
  \bibfield  {author} {\bibinfo {author} {\bibfnamefont {A.}~\bibnamefont
  {Hayward}}, \bibinfo {author} {\bibfnamefont {C.}~\bibnamefont {Schweizer}},
  \bibinfo {author} {\bibfnamefont {M.}~\bibnamefont {Lohse}}, \bibinfo
  {author} {\bibfnamefont {M.}~\bibnamefont {Aidelsburger}}, \ and\ \bibinfo
  {author} {\bibfnamefont {F.}~\bibnamefont {Heidrich-Meisner}},\ }\href
  {\doibase 10.1103/PhysRevB.98.245148} {\bibfield  {journal} {\bibinfo
  {journal} {Phys. Rev. B}\ }\textbf {\bibinfo {volume} {98}},\ \bibinfo
  {pages} {245148} (\bibinfo {year} {2018})}\BibitemShut {NoStop}%
\bibitem [{sup()}]{supmat}%
  \BibitemOpen
  \href@noop {} {\emph {\bibinfo {title} {{\rm See {Supplementary Material},
  which includes Refs.~\cite{Schulz1986, kitazawa1996phase,
  kitazawa1997critical, ejima2018exotic, Berg2008, GiamarchiBook,
  Takayoshi2018, Dhar2012, Greschner2013, Mishra12, Dalmonte14}, for
  details.}}}}\BibitemShut {Stop}%
\bibitem [{\citenamefont {Zak}(1989)}]{Zak1989}%
  \BibitemOpen
  \bibfield  {author} {\bibinfo {author} {\bibfnamefont {J.}~\bibnamefont
  {Zak}},\ }\href {\doibase 10.1103/PhysRevLett.62.2747} {\bibfield  {journal}
  {\bibinfo  {journal} {Phys. Rev. Lett.}\ }\textbf {\bibinfo {volume} {62}},\
  \bibinfo {pages} {2747} (\bibinfo {year} {1989})}\BibitemShut {NoStop}%
\bibitem [{\citenamefont {Mondal}\ \emph {et~al.}(2019)\citenamefont {Mondal},
  \citenamefont {Greschner},\ and\ \citenamefont {Mishra}}]{Mondal2019}%
  \BibitemOpen
  \bibfield  {author} {\bibinfo {author} {\bibfnamefont {S.}~\bibnamefont
  {Mondal}}, \bibinfo {author} {\bibfnamefont {S.}~\bibnamefont {Greschner}}, \
  and\ \bibinfo {author} {\bibfnamefont {T.}~\bibnamefont {Mishra}},\ }\href
  {\doibase 10.1103/PhysRevA.100.013627} {\bibfield  {journal} {\bibinfo
  {journal} {Phys. Rev. A}\ }\textbf {\bibinfo {volume} {100}},\ \bibinfo
  {pages} {013627} (\bibinfo {year} {2019})}\BibitemShut {NoStop}%
\bibitem [{\citenamefont {Daley}\ \emph
  {et~al.}(2009{\natexlab{a}})\citenamefont {Daley}, \citenamefont {Taylor},
  \citenamefont {Diehl}, \citenamefont {Baranov},\ and\ \citenamefont
  {Zoller}}]{Daley2009}%
  \BibitemOpen
  \bibfield  {author} {\bibinfo {author} {\bibfnamefont {A.~J.}\ \bibnamefont
  {Daley}}, \bibinfo {author} {\bibfnamefont {J.~M.}\ \bibnamefont {Taylor}},
  \bibinfo {author} {\bibfnamefont {S.}~\bibnamefont {Diehl}}, \bibinfo
  {author} {\bibfnamefont {M.}~\bibnamefont {Baranov}}, \ and\ \bibinfo
  {author} {\bibfnamefont {P.}~\bibnamefont {Zoller}},\ }\href {\doibase
  10.1103/PhysRevLett.102.040402} {\bibfield  {journal} {\bibinfo  {journal}
  {Phys. Rev. Lett.}\ }\textbf {\bibinfo {volume} {102}},\ \bibinfo {pages}
  {040402} (\bibinfo {year} {2009}{\natexlab{a}})}\BibitemShut {NoStop}%
\bibitem [{Bac()}]{Backgroundnote}%
  \BibitemOpen
  \href@noop {} {\emph {\bibinfo {title} {{\rm In order to visualize the
  properties of the edges clearly we subtract the background average density
  $\bar{n}_j = \la n_{L/2 + (j\, {\rm mod}\, 2)} \ra$}}}}\BibitemShut {NoStop}%
\bibitem [{\citenamefont {Kane}\ and\ \citenamefont {Mele}(2005)}]{kane2005}%
  \BibitemOpen
  \bibfield  {author} {\bibinfo {author} {\bibfnamefont {C.~L.}\ \bibnamefont
  {Kane}}\ and\ \bibinfo {author} {\bibfnamefont {E.~J.}\ \bibnamefont
  {Mele}},\ }\href@noop {} {\bibfield  {journal} {\bibinfo  {journal} {Physical
  review letters}\ }\textbf {\bibinfo {volume} {95}},\ \bibinfo {pages}
  {146802} (\bibinfo {year} {2005})}\BibitemShut {NoStop}%
\bibitem [{\citenamefont {Tsukano}\ and\ \citenamefont
  {Nomura}(1998{\natexlab{a}})}]{tsukano1998berezinskii}%
  \BibitemOpen
  \bibfield  {author} {\bibinfo {author} {\bibfnamefont {M.}~\bibnamefont
  {Tsukano}}\ and\ \bibinfo {author} {\bibfnamefont {K.}~\bibnamefont
  {Nomura}},\ }\href@noop {} {\bibfield  {journal} {\bibinfo  {journal}
  {Journal of the Physical Society of Japan}\ }\textbf {\bibinfo {volume}
  {67}},\ \bibinfo {pages} {302} (\bibinfo {year}
  {1998}{\natexlab{a}})}\BibitemShut {NoStop}%
\bibitem [{\citenamefont {Tsukano}\ and\ \citenamefont
  {Nomura}(1998{\natexlab{b}})}]{tsukano1998spin}%
  \BibitemOpen
  \bibfield  {author} {\bibinfo {author} {\bibfnamefont {M.}~\bibnamefont
  {Tsukano}}\ and\ \bibinfo {author} {\bibfnamefont {K.}~\bibnamefont
  {Nomura}},\ }\href@noop {} {\bibfield  {journal} {\bibinfo  {journal}
  {Physical Review B}\ }\textbf {\bibinfo {volume} {57}},\ \bibinfo {pages}
  {R8087} (\bibinfo {year} {1998}{\natexlab{b}})}\BibitemShut {NoStop}%
\bibitem [{\citenamefont {Daley}\ \emph
  {et~al.}(2009{\natexlab{b}})\citenamefont {Daley}, \citenamefont {Taylor},
  \citenamefont {Diehl}, \citenamefont {Baranov},\ and\ \citenamefont
  {Zoller}}]{baranovprl}%
  \BibitemOpen
  \bibfield  {author} {\bibinfo {author} {\bibfnamefont {A.~J.}\ \bibnamefont
  {Daley}}, \bibinfo {author} {\bibfnamefont {J.~M.}\ \bibnamefont {Taylor}},
  \bibinfo {author} {\bibfnamefont {S.}~\bibnamefont {Diehl}}, \bibinfo
  {author} {\bibfnamefont {M.}~\bibnamefont {Baranov}}, \ and\ \bibinfo
  {author} {\bibfnamefont {P.}~\bibnamefont {Zoller}},\ }\href {\doibase
  10.1103/PhysRevLett.102.040402} {\bibfield  {journal} {\bibinfo  {journal}
  {Phys. Rev. Lett.}\ }\textbf {\bibinfo {volume} {102}},\ \bibinfo {pages}
  {040402} (\bibinfo {year} {2009}{\natexlab{b}})}\BibitemShut {NoStop}%
\bibitem [{\citenamefont {Petrov}(2014{\natexlab{a}})}]{petrov1}%
  \BibitemOpen
  \bibfield  {author} {\bibinfo {author} {\bibfnamefont {D.~S.}\ \bibnamefont
  {Petrov}},\ }\href {\doibase 10.1103/PhysRevLett.112.103201} {\bibfield
  {journal} {\bibinfo  {journal} {Phys. Rev. Lett.}\ }\textbf {\bibinfo
  {volume} {112}},\ \bibinfo {pages} {103201} (\bibinfo {year}
  {2014}{\natexlab{a}})}\BibitemShut {NoStop}%
\bibitem [{\citenamefont {Petrov}(2014{\natexlab{b}})}]{petrov2}%
  \BibitemOpen
  \bibfield  {author} {\bibinfo {author} {\bibfnamefont {D.~S.}\ \bibnamefont
  {Petrov}},\ }\href {\doibase 10.1103/PhysRevA.90.021601} {\bibfield
  {journal} {\bibinfo  {journal} {Phys. Rev. A}\ }\textbf {\bibinfo {volume}
  {90}},\ \bibinfo {pages} {021601} (\bibinfo {year}
  {2014}{\natexlab{b}})}\BibitemShut {NoStop}%
\bibitem [{\citenamefont {Johnson}\ \emph {et~al.}(2009)\citenamefont
  {Johnson}, \citenamefont {Tiesinga}, \citenamefont {Porto},\ and\
  \citenamefont {Williams}}]{tiesinga}%
  \BibitemOpen
  \bibfield  {author} {\bibinfo {author} {\bibfnamefont {P.~R.}\ \bibnamefont
  {Johnson}}, \bibinfo {author} {\bibfnamefont {E.}~\bibnamefont {Tiesinga}},
  \bibinfo {author} {\bibfnamefont {J.~V.}\ \bibnamefont {Porto}}, \ and\
  \bibinfo {author} {\bibfnamefont {C.~J.}\ \bibnamefont {Williams}},\
  }\href@noop {} {\bibfield  {journal} {\bibinfo  {journal} {New Journal of
  Physics}\ }\textbf {\bibinfo {volume} {11}},\ \bibinfo {pages} {093022}
  (\bibinfo {year} {2009})}\BibitemShut {NoStop}%
\bibitem [{\citenamefont {Safavi-Naini}\ \emph {et~al.}(2012)\citenamefont
  {Safavi-Naini}, \citenamefont {von Stecher}, \citenamefont
  {Capogrosso-Sansone},\ and\ \citenamefont {Rittenhouse}}]{sansone}%
  \BibitemOpen
  \bibfield  {author} {\bibinfo {author} {\bibfnamefont {A.}~\bibnamefont
  {Safavi-Naini}}, \bibinfo {author} {\bibfnamefont {J.}~\bibnamefont {von
  Stecher}}, \bibinfo {author} {\bibfnamefont {B.}~\bibnamefont
  {Capogrosso-Sansone}}, \ and\ \bibinfo {author} {\bibfnamefont {S.~T.}\
  \bibnamefont {Rittenhouse}},\ }\href {\doibase
  10.1103/PhysRevLett.109.135302} {\bibfield  {journal} {\bibinfo  {journal}
  {Phys. Rev. Lett.}\ }\textbf {\bibinfo {volume} {109}},\ \bibinfo {pages}
  {135302} (\bibinfo {year} {2012})}\BibitemShut {NoStop}%
\bibitem [{\citenamefont {Greschner}\ and\ \citenamefont
  {Santos}(2015)}]{greschner2015anyon}%
  \BibitemOpen
  \bibfield  {author} {\bibinfo {author} {\bibfnamefont {S.}~\bibnamefont
  {Greschner}}\ and\ \bibinfo {author} {\bibfnamefont {L.}~\bibnamefont
  {Santos}},\ }\href@noop {} {\bibfield  {journal} {\bibinfo  {journal}
  {Physical review letters}\ }\textbf {\bibinfo {volume} {115}},\ \bibinfo
  {pages} {053002} (\bibinfo {year} {2015})}\BibitemShut {NoStop}%
\bibitem [{\citenamefont {Yang}\ \emph {et~al.}(2019)\citenamefont {Yang},
  \citenamefont {Sun}, \citenamefont {Huang}, \citenamefont {Wang},
  \citenamefont {Deng}, \citenamefont {Dai}, \citenamefont {Yuan},\ and\
  \citenamefont {Pan}}]{Bing2019staggered}%
  \BibitemOpen
  \bibfield  {author} {\bibinfo {author} {\bibfnamefont {B.}~\bibnamefont
  {Yang}}, \bibinfo {author} {\bibfnamefont {H.}~\bibnamefont {Sun}}, \bibinfo
  {author} {\bibfnamefont {C.-J.}\ \bibnamefont {Huang}}, \bibinfo {author}
  {\bibfnamefont {H.-Y.}\ \bibnamefont {Wang}}, \bibinfo {author}
  {\bibfnamefont {Y.-J.}\ \bibnamefont {Deng}}, \bibinfo {author}
  {\bibfnamefont {H.-N.}\ \bibnamefont {Dai}}, \bibinfo {author} {\bibfnamefont
  {Z.-S.}\ \bibnamefont {Yuan}}, \ and\ \bibinfo {author} {\bibfnamefont
  {J.-W.}\ \bibnamefont {Pan}},\ }\href@noop {} {\bibfield  {journal} {\bibinfo
   {journal} {arXiv preprint arXiv:1901.01146}\ } (\bibinfo {year}
  {2019})}\BibitemShut {NoStop}%
\bibitem [{\citenamefont {Winkler}\ \emph {et~al.}(2006)\citenamefont
  {Winkler}, \citenamefont {Thalhammer}, \citenamefont {Lang}, \citenamefont
  {Grimm}, \citenamefont {Denschlag}, \citenamefont {Daley}, \citenamefont
  {Kantian}, \citenamefont {B{\"u}chler},\ and\ \citenamefont
  {Zoller}}]{winkler2006repulsively}%
  \BibitemOpen
  \bibfield  {author} {\bibinfo {author} {\bibfnamefont {K.}~\bibnamefont
  {Winkler}}, \bibinfo {author} {\bibfnamefont {G.}~\bibnamefont {Thalhammer}},
  \bibinfo {author} {\bibfnamefont {F.}~\bibnamefont {Lang}}, \bibinfo {author}
  {\bibfnamefont {R.}~\bibnamefont {Grimm}}, \bibinfo {author} {\bibfnamefont
  {J.~H.}\ \bibnamefont {Denschlag}}, \bibinfo {author} {\bibfnamefont
  {A.}~\bibnamefont {Daley}}, \bibinfo {author} {\bibfnamefont
  {A.}~\bibnamefont {Kantian}}, \bibinfo {author} {\bibfnamefont
  {H.}~\bibnamefont {B{\"u}chler}}, \ and\ \bibinfo {author} {\bibfnamefont
  {P.}~\bibnamefont {Zoller}},\ }\href@noop {} {\bibfield  {journal} {\bibinfo
  {journal} {Nature}\ }\textbf {\bibinfo {volume} {441}},\ \bibinfo {pages}
  {853} (\bibinfo {year} {2006})}\BibitemShut {NoStop}%
\bibitem [{\citenamefont {Petrosyan}\ \emph {et~al.}(2007)\citenamefont
  {Petrosyan}, \citenamefont {Schmidt}, \citenamefont {Anglin},\ and\
  \citenamefont {Fleischhauer}}]{petrosyan2007quantum}%
  \BibitemOpen
  \bibfield  {author} {\bibinfo {author} {\bibfnamefont {D.}~\bibnamefont
  {Petrosyan}}, \bibinfo {author} {\bibfnamefont {B.}~\bibnamefont {Schmidt}},
  \bibinfo {author} {\bibfnamefont {J.~R.}\ \bibnamefont {Anglin}}, \ and\
  \bibinfo {author} {\bibfnamefont {M.}~\bibnamefont {Fleischhauer}},\
  }\href@noop {} {\bibfield  {journal} {\bibinfo  {journal} {Physical Review
  A}\ }\textbf {\bibinfo {volume} {76}},\ \bibinfo {pages} {033606} (\bibinfo
  {year} {2007})}\BibitemShut {NoStop}%
\bibitem [{\citenamefont {{Will Sebastian}}\ \emph {et~al.}(2010)\citenamefont
  {{Will Sebastian}}, \citenamefont {{Best Thorsten}}, \citenamefont
  {{Schneider Ulrich}}, \citenamefont {{Hackerm{\"u}ller Lucia}}, \citenamefont
  {{L{\"u}hmann Dirk-S{\"o}ren}},\ and\ \citenamefont {{Bloch
  Immanuel}}}]{will2010}%
  \BibitemOpen
  \bibfield  {author} {\bibinfo {author} {\bibnamefont {{Will Sebastian}}},
  \bibinfo {author} {\bibnamefont {{Best Thorsten}}}, \bibinfo {author}
  {\bibnamefont {{Schneider Ulrich}}}, \bibinfo {author} {\bibnamefont
  {{Hackerm{\"u}ller Lucia}}}, \bibinfo {author} {\bibnamefont {{L{\"u}hmann
  Dirk-S{\"o}ren}}}, \ and\ \bibinfo {author} {\bibnamefont {{Bloch
  Immanuel}}},\ }\href {\doibase https://doi.org/10.1038/nature09036
  10.1038/nature09036} {\bibfield  {journal} {\bibinfo  {journal} {Nature}\
  }\textbf {\bibinfo {volume} {465}},\ \bibinfo {pages} {197} (\bibinfo {year}
  {2010})}\BibitemShut {NoStop}%
\bibitem [{\citenamefont {Schulz}(1986)}]{Schulz1986}%
  \BibitemOpen
  \bibfield  {author} {\bibinfo {author} {\bibfnamefont {H.~J.}\ \bibnamefont
  {Schulz}},\ }\href {\doibase 10.1103/PhysRevB.34.6372} {\bibfield  {journal}
  {\bibinfo  {journal} {Phys. Rev. B}\ }\textbf {\bibinfo {volume} {34}},\
  \bibinfo {pages} {6372} (\bibinfo {year} {1986})}\BibitemShut {NoStop}%
\bibitem [{\citenamefont {Kitazawa}\ \emph {et~al.}(1996)\citenamefont
  {Kitazawa}, \citenamefont {Nomura},\ and\ \citenamefont
  {Okamoto}}]{kitazawa1996phase}%
  \BibitemOpen
  \bibfield  {author} {\bibinfo {author} {\bibfnamefont {A.}~\bibnamefont
  {Kitazawa}}, \bibinfo {author} {\bibfnamefont {K.}~\bibnamefont {Nomura}}, \
  and\ \bibinfo {author} {\bibfnamefont {K.}~\bibnamefont {Okamoto}},\
  }\href@noop {} {\bibfield  {journal} {\bibinfo  {journal} {Physical review
  letters}\ }\textbf {\bibinfo {volume} {76}},\ \bibinfo {pages} {4038}
  (\bibinfo {year} {1996})}\BibitemShut {NoStop}%
\bibitem [{\citenamefont {Kitazawa}\ and\ \citenamefont
  {Nomura}(1997)}]{kitazawa1997critical}%
  \BibitemOpen
  \bibfield  {author} {\bibinfo {author} {\bibfnamefont {A.}~\bibnamefont
  {Kitazawa}}\ and\ \bibinfo {author} {\bibfnamefont {K.}~\bibnamefont
  {Nomura}},\ }\href@noop {} {\bibfield  {journal} {\bibinfo  {journal}
  {Journal of the Physical Society of Japan}\ }\textbf {\bibinfo {volume}
  {66}},\ \bibinfo {pages} {3944} (\bibinfo {year} {1997})}\BibitemShut
  {NoStop}%
\bibitem [{\citenamefont {Ejima}\ \emph {et~al.}(2018)\citenamefont {Ejima},
  \citenamefont {Yamaguchi}, \citenamefont {Essler}, \citenamefont {Lange},
  \citenamefont {Ohta},\ and\ \citenamefont {Fehske}}]{ejima2018exotic}%
  \BibitemOpen
  \bibfield  {author} {\bibinfo {author} {\bibfnamefont {S.}~\bibnamefont
  {Ejima}}, \bibinfo {author} {\bibfnamefont {T.}~\bibnamefont {Yamaguchi}},
  \bibinfo {author} {\bibfnamefont {F.}~\bibnamefont {Essler}}, \bibinfo
  {author} {\bibfnamefont {F.}~\bibnamefont {Lange}}, \bibinfo {author}
  {\bibfnamefont {Y.}~\bibnamefont {Ohta}}, \ and\ \bibinfo {author}
  {\bibfnamefont {H.}~\bibnamefont {Fehske}},\ }\href@noop {} {\bibfield
  {journal} {\bibinfo  {journal} {SciPost Physics}\ }\textbf {\bibinfo {volume}
  {5}} (\bibinfo {year} {2018})}\BibitemShut {NoStop}%
\bibitem [{\citenamefont {Berg}\ \emph {et~al.}(2008)\citenamefont {Berg},
  \citenamefont {Dalla~Torre}, \citenamefont {Giamarchi},\ and\ \citenamefont
  {Altman}}]{Berg2008}%
  \BibitemOpen
  \bibfield  {author} {\bibinfo {author} {\bibfnamefont {E.}~\bibnamefont
  {Berg}}, \bibinfo {author} {\bibfnamefont {E.~G.}\ \bibnamefont
  {Dalla~Torre}}, \bibinfo {author} {\bibfnamefont {T.}~\bibnamefont
  {Giamarchi}}, \ and\ \bibinfo {author} {\bibfnamefont {E.}~\bibnamefont
  {Altman}},\ }\href@noop {} {\bibfield  {journal} {\bibinfo  {journal}
  {Physical Review B}\ }\textbf {\bibinfo {volume} {77}},\ \bibinfo {pages}
  {245119} (\bibinfo {year} {2008})}\BibitemShut {NoStop}%
\bibitem [{\citenamefont {Giamarchi}(2003)}]{GiamarchiBook}%
  \BibitemOpen
  \bibfield  {author} {\bibinfo {author} {\bibfnamefont {T.}~\bibnamefont
  {Giamarchi}},\ }\href@noop {} {\emph {\bibinfo {title} {Quantum physics in
  one dimension}}},\ Vol.\ \bibinfo {volume} {121}\ (\bibinfo  {publisher}
  {Clarendon press},\ \bibinfo {year} {2003})\BibitemShut {NoStop}%
\bibitem [{\citenamefont {Takayoshi}\ \emph {et~al.}(2018)\citenamefont
  {Takayoshi}, \citenamefont {Furuya},\ and\ \citenamefont
  {Giamarchi}}]{Takayoshi2018}%
  \BibitemOpen
  \bibfield  {author} {\bibinfo {author} {\bibfnamefont {S.}~\bibnamefont
  {Takayoshi}}, \bibinfo {author} {\bibfnamefont {S.~C.}\ \bibnamefont
  {Furuya}}, \ and\ \bibinfo {author} {\bibfnamefont {T.}~\bibnamefont
  {Giamarchi}},\ }\href {\doibase 10.1103/PhysRevB.98.184429} {\bibfield
  {journal} {\bibinfo  {journal} {Phys. Rev. B}\ }\textbf {\bibinfo {volume}
  {98}},\ \bibinfo {pages} {184429} (\bibinfo {year} {2018})}\BibitemShut
  {NoStop}%
\bibitem [{\citenamefont {Dhar}\ \emph {et~al.}(2012)\citenamefont {Dhar},
  \citenamefont {Maji}, \citenamefont {Mishra}, \citenamefont {Pai},
  \citenamefont {Mukerjee},\ and\ \citenamefont {Paramekanti}}]{Dhar2012}%
  \BibitemOpen
  \bibfield  {author} {\bibinfo {author} {\bibfnamefont {A.}~\bibnamefont
  {Dhar}}, \bibinfo {author} {\bibfnamefont {M.}~\bibnamefont {Maji}}, \bibinfo
  {author} {\bibfnamefont {T.}~\bibnamefont {Mishra}}, \bibinfo {author}
  {\bibfnamefont {R.~V.}\ \bibnamefont {Pai}}, \bibinfo {author} {\bibfnamefont
  {S.}~\bibnamefont {Mukerjee}}, \ and\ \bibinfo {author} {\bibfnamefont
  {A.}~\bibnamefont {Paramekanti}},\ }\href {\doibase
  10.1103/PhysRevA.85.041602} {\bibfield  {journal} {\bibinfo  {journal} {Phys.
  Rev. A}\ }\textbf {\bibinfo {volume} {85}},\ \bibinfo {pages} {041602}
  (\bibinfo {year} {2012})}\BibitemShut {NoStop}%
\bibitem [{\citenamefont {Greschner}\ \emph {et~al.}(2013)\citenamefont
  {Greschner}, \citenamefont {Santos},\ and\ \citenamefont
  {Vekua}}]{Greschner2013}%
  \BibitemOpen
  \bibfield  {author} {\bibinfo {author} {\bibfnamefont {S.}~\bibnamefont
  {Greschner}}, \bibinfo {author} {\bibfnamefont {L.}~\bibnamefont {Santos}}, \
  and\ \bibinfo {author} {\bibfnamefont {T.}~\bibnamefont {Vekua}},\ }\href
  {\doibase 10.1103/PhysRevA.87.033609} {\bibfield  {journal} {\bibinfo
  {journal} {Phys. Rev. A}\ }\textbf {\bibinfo {volume} {87}},\ \bibinfo
  {pages} {033609} (\bibinfo {year} {2013})}\BibitemShut {NoStop}%
\bibitem [{\citenamefont {Mishra}\ \emph {et~al.}(2011)\citenamefont {Mishra},
  \citenamefont {Carrasquilla},\ and\ \citenamefont {Rigol}}]{Mishra12}%
  \BibitemOpen
  \bibfield  {author} {\bibinfo {author} {\bibfnamefont {T.}~\bibnamefont
  {Mishra}}, \bibinfo {author} {\bibfnamefont {J.}~\bibnamefont
  {Carrasquilla}}, \ and\ \bibinfo {author} {\bibfnamefont {M.}~\bibnamefont
  {Rigol}},\ }\href {\doibase 10.1103/PhysRevB.84.115135} {\bibfield  {journal}
  {\bibinfo  {journal} {Phys. Rev. B}\ }\textbf {\bibinfo {volume} {84}},\
  \bibinfo {pages} {115135} (\bibinfo {year} {2011})}\BibitemShut {NoStop}%
\bibitem [{\citenamefont {Dalmonte}\ \emph {et~al.}(2015)\citenamefont
  {Dalmonte}, \citenamefont {Carrasquilla}, \citenamefont {Taddia},
  \citenamefont {Ercolessi},\ and\ \citenamefont {Rigol}}]{Dalmonte14}%
  \BibitemOpen
  \bibfield  {author} {\bibinfo {author} {\bibfnamefont {M.}~\bibnamefont
  {Dalmonte}}, \bibinfo {author} {\bibfnamefont {J.}~\bibnamefont
  {Carrasquilla}}, \bibinfo {author} {\bibfnamefont {L.}~\bibnamefont
  {Taddia}}, \bibinfo {author} {\bibfnamefont {E.}~\bibnamefont {Ercolessi}}, \
  and\ \bibinfo {author} {\bibfnamefont {M.}~\bibnamefont {Rigol}},\ }\href
  {\doibase 10.1103/PhysRevB.91.165136} {\bibfield  {journal} {\bibinfo
  {journal} {Phys. Rev. B}\ }\textbf {\bibinfo {volume} {91}},\ \bibinfo
  {pages} {165136} (\bibinfo {year} {2015})}\BibitemShut {NoStop}%
\end{thebibliography}%


\begin{thebibliography}{16}%
\makeatletter
\providecommand \@ifxundefined [1]{%
 \@ifx{#1\undefined}
}%
\providecommand \@ifnum [1]{%
 \ifnum #1\expandafter \@firstoftwo
 \else \expandafter \@secondoftwo
 \fi
}%
\providecommand \@ifx [1]{%
 \ifx #1\expandafter \@firstoftwo
 \else \expandafter \@secondoftwo
 \fi
}%
\providecommand \natexlab [1]{#1}%
\providecommand \enquote  [1]{``#1''}%
\providecommand \bibnamefont  [1]{#1}%
\providecommand \bibfnamefont [1]{#1}%
\providecommand \citenamefont [1]{#1}%
\providecommand \href@noop [0]{\@secondoftwo}%
\providecommand \href [0]{\begingroup \@sanitize@url \@href}%
\providecommand \@href[1]{\@@startlink{#1}\@@href}%
\providecommand \@@href[1]{\endgroup#1\@@endlink}%
\providecommand \@sanitize@url [0]{\catcode `\\12\catcode `\$12\catcode
  `\&12\catcode `\#12\catcode `\^12\catcode `\_12\catcode `\%12\relax}%
\providecommand \@@startlink[1]{}%
\providecommand \@@endlink[0]{}%
\providecommand \url  [0]{\begingroup\@sanitize@url \@url }%
\providecommand \@url [1]{\endgroup\@href {#1}{\urlprefix }}%
\providecommand \urlprefix  [0]{URL }%
\providecommand \Eprint [0]{\href }%
\providecommand \doibase [0]{http://dx.doi.org/}%
\providecommand \selectlanguage [0]{\@gobble}%
\providecommand \bibinfo  [0]{\@secondoftwo}%
\providecommand \bibfield  [0]{\@secondoftwo}%
\providecommand \translation [1]{[#1]}%
\providecommand \BibitemOpen [0]{}%
\providecommand \bibitemStop [0]{}%
\providecommand \bibitemNoStop [0]{.\EOS\space}%
\providecommand \EOS [0]{\spacefactor3000\relax}%
\providecommand \BibitemShut  [1]{\csname bibitem#1\endcsname}%
\let\auto@bib@innerbib\@empty
\bibitem [{\citenamefont {Mondal}\ \emph {et~al.}(2019)\citenamefont {Mondal},
  \citenamefont {Greschner},\ and\ \citenamefont {Mishra}}]{Mondal2019}%
  \BibitemOpen
  \bibfield  {author} {\bibinfo {author} {\bibfnamefont {S.}~\bibnamefont
  {Mondal}}, \bibinfo {author} {\bibfnamefont {S.}~\bibnamefont {Greschner}}, \
  and\ \bibinfo {author} {\bibfnamefont {T.}~\bibnamefont {Mishra}},\ }\href
  {\doibase 10.1103/PhysRevA.100.013627} {\bibfield  {journal} {\bibinfo
  {journal} {Phys. Rev. A}\ }\textbf {\bibinfo {volume} {100}},\ \bibinfo
  {pages} {013627} (\bibinfo {year} {2019})}\BibitemShut {NoStop}%
\bibitem [{\citenamefont {Schulz}(1986)}]{Schulz1986}%
  \BibitemOpen
  \bibfield  {author} {\bibinfo {author} {\bibfnamefont {H.~J.}\ \bibnamefont
  {Schulz}},\ }\href {\doibase 10.1103/PhysRevB.34.6372} {\bibfield  {journal}
  {\bibinfo  {journal} {Phys. Rev. B}\ }\textbf {\bibinfo {volume} {34}},\
  \bibinfo {pages} {6372} (\bibinfo {year} {1986})}\BibitemShut {NoStop}%
\bibitem [{\citenamefont {Kitazawa}\ \emph {et~al.}(1996)\citenamefont
  {Kitazawa}, \citenamefont {Nomura},\ and\ \citenamefont
  {Okamoto}}]{kitazawa1996phase}%
  \BibitemOpen
  \bibfield  {author} {\bibinfo {author} {\bibfnamefont {A.}~\bibnamefont
  {Kitazawa}}, \bibinfo {author} {\bibfnamefont {K.}~\bibnamefont {Nomura}}, \
  and\ \bibinfo {author} {\bibfnamefont {K.}~\bibnamefont {Okamoto}},\
  }\href@noop {} {\bibfield  {journal} {\bibinfo  {journal} {Physical review
  letters}\ }\textbf {\bibinfo {volume} {76}},\ \bibinfo {pages} {4038}
  (\bibinfo {year} {1996})}\BibitemShut {NoStop}%
\bibitem [{\citenamefont {Kitazawa}\ and\ \citenamefont
  {Nomura}(1997)}]{kitazawa1997critical}%
  \BibitemOpen
  \bibfield  {author} {\bibinfo {author} {\bibfnamefont {A.}~\bibnamefont
  {Kitazawa}}\ and\ \bibinfo {author} {\bibfnamefont {K.}~\bibnamefont
  {Nomura}},\ }\href@noop {} {\bibfield  {journal} {\bibinfo  {journal}
  {Journal of the Physical Society of Japan}\ }\textbf {\bibinfo {volume}
  {66}},\ \bibinfo {pages} {3944} (\bibinfo {year} {1997})}\BibitemShut
  {NoStop}%
\bibitem [{\citenamefont {Ejima}\ \emph {et~al.}(2018)\citenamefont {Ejima},
  \citenamefont {Yamaguchi}, \citenamefont {Essler}, \citenamefont {Lange},
  \citenamefont {Ohta},\ and\ \citenamefont {Fehske}}]{ejima2018exotic}%
  \BibitemOpen
  \bibfield  {author} {\bibinfo {author} {\bibfnamefont {S.}~\bibnamefont
  {Ejima}}, \bibinfo {author} {\bibfnamefont {T.}~\bibnamefont {Yamaguchi}},
  \bibinfo {author} {\bibfnamefont {F.}~\bibnamefont {Essler}}, \bibinfo
  {author} {\bibfnamefont {F.}~\bibnamefont {Lange}}, \bibinfo {author}
  {\bibfnamefont {Y.}~\bibnamefont {Ohta}}, \ and\ \bibinfo {author}
  {\bibfnamefont {H.}~\bibnamefont {Fehske}},\ }\href@noop {} {\bibfield
  {journal} {\bibinfo  {journal} {SciPost Physics}\ }\textbf {\bibinfo {volume}
  {5}} (\bibinfo {year} {2018})}\BibitemShut {NoStop}%
\bibitem [{\citenamefont {Berg}\ \emph {et~al.}(2008)\citenamefont {Berg},
  \citenamefont {Dalla~Torre}, \citenamefont {Giamarchi},\ and\ \citenamefont
  {Altman}}]{Berg2008}%
  \BibitemOpen
  \bibfield  {author} {\bibinfo {author} {\bibfnamefont {E.}~\bibnamefont
  {Berg}}, \bibinfo {author} {\bibfnamefont {E.~G.}\ \bibnamefont
  {Dalla~Torre}}, \bibinfo {author} {\bibfnamefont {T.}~\bibnamefont
  {Giamarchi}}, \ and\ \bibinfo {author} {\bibfnamefont {E.}~\bibnamefont
  {Altman}},\ }\href@noop {} {\bibfield  {journal} {\bibinfo  {journal}
  {Physical Review B}\ }\textbf {\bibinfo {volume} {77}},\ \bibinfo {pages}
  {245119} (\bibinfo {year} {2008})}\BibitemShut {NoStop}%
\bibitem [{\citenamefont {Giamarchi}(2003)}]{GiamarchiBook}%
  \BibitemOpen
  \bibfield  {author} {\bibinfo {author} {\bibfnamefont {T.}~\bibnamefont
  {Giamarchi}},\ }\href@noop {} {\emph {\bibinfo {title} {Quantum physics in
  one dimension}}},\ Vol.\ \bibinfo {volume} {121}\ (\bibinfo  {publisher}
  {Clarendon press},\ \bibinfo {year} {2003})\BibitemShut {NoStop}%
\bibitem [{\citenamefont {Takayoshi}\ \emph {et~al.}(2018)\citenamefont
  {Takayoshi}, \citenamefont {Furuya},\ and\ \citenamefont
  {Giamarchi}}]{Takayoshi2018}%
  \BibitemOpen
  \bibfield  {author} {\bibinfo {author} {\bibfnamefont {S.}~\bibnamefont
  {Takayoshi}}, \bibinfo {author} {\bibfnamefont {S.~C.}\ \bibnamefont
  {Furuya}}, \ and\ \bibinfo {author} {\bibfnamefont {T.}~\bibnamefont
  {Giamarchi}},\ }\href {\doibase 10.1103/PhysRevB.98.184429} {\bibfield
  {journal} {\bibinfo  {journal} {Phys. Rev. B}\ }\textbf {\bibinfo {volume}
  {98}},\ \bibinfo {pages} {184429} (\bibinfo {year} {2018})}\BibitemShut
  {NoStop}%
\bibitem [{\citenamefont {Dhar}\ \emph {et~al.}(2012)\citenamefont {Dhar},
  \citenamefont {Maji}, \citenamefont {Mishra}, \citenamefont {Pai},
  \citenamefont {Mukerjee},\ and\ \citenamefont {Paramekanti}}]{Dhar2012}%
  \BibitemOpen
  \bibfield  {author} {\bibinfo {author} {\bibfnamefont {A.}~\bibnamefont
  {Dhar}}, \bibinfo {author} {\bibfnamefont {M.}~\bibnamefont {Maji}}, \bibinfo
  {author} {\bibfnamefont {T.}~\bibnamefont {Mishra}}, \bibinfo {author}
  {\bibfnamefont {R.~V.}\ \bibnamefont {Pai}}, \bibinfo {author} {\bibfnamefont
  {S.}~\bibnamefont {Mukerjee}}, \ and\ \bibinfo {author} {\bibfnamefont
  {A.}~\bibnamefont {Paramekanti}},\ }\href {\doibase
  10.1103/PhysRevA.85.041602} {\bibfield  {journal} {\bibinfo  {journal} {Phys.
  Rev. A}\ }\textbf {\bibinfo {volume} {85}},\ \bibinfo {pages} {041602}
  (\bibinfo {year} {2012})}\BibitemShut {NoStop}%
\bibitem [{\citenamefont {Greschner}\ \emph {et~al.}(2013)\citenamefont
  {Greschner}, \citenamefont {Santos},\ and\ \citenamefont
  {Vekua}}]{Greschner2013}%
  \BibitemOpen
  \bibfield  {author} {\bibinfo {author} {\bibfnamefont {S.}~\bibnamefont
  {Greschner}}, \bibinfo {author} {\bibfnamefont {L.}~\bibnamefont {Santos}}, \
  and\ \bibinfo {author} {\bibfnamefont {T.}~\bibnamefont {Vekua}},\ }\href
  {\doibase 10.1103/PhysRevA.87.033609} {\bibfield  {journal} {\bibinfo
  {journal} {Phys. Rev. A}\ }\textbf {\bibinfo {volume} {87}},\ \bibinfo
  {pages} {033609} (\bibinfo {year} {2013})}\BibitemShut {NoStop}%
\bibitem [{\citenamefont {Mishra}\ \emph {et~al.}(2011)\citenamefont {Mishra},
  \citenamefont {Carrasquilla},\ and\ \citenamefont {Rigol}}]{Mishra12}%
  \BibitemOpen
  \bibfield  {author} {\bibinfo {author} {\bibfnamefont {T.}~\bibnamefont
  {Mishra}}, \bibinfo {author} {\bibfnamefont {J.}~\bibnamefont
  {Carrasquilla}}, \ and\ \bibinfo {author} {\bibfnamefont {M.}~\bibnamefont
  {Rigol}},\ }\href {\doibase 10.1103/PhysRevB.84.115135} {\bibfield  {journal}
  {\bibinfo  {journal} {Phys. Rev. B}\ }\textbf {\bibinfo {volume} {84}},\
  \bibinfo {pages} {115135} (\bibinfo {year} {2011})}\BibitemShut {NoStop}%
\bibitem [{\citenamefont {Dalmonte}\ \emph {et~al.}(2015)\citenamefont
  {Dalmonte}, \citenamefont {Carrasquilla}, \citenamefont {Taddia},
  \citenamefont {Ercolessi},\ and\ \citenamefont {Rigol}}]{Dalmonte14}%
  \BibitemOpen
  \bibfield  {author} {\bibinfo {author} {\bibfnamefont {M.}~\bibnamefont
  {Dalmonte}}, \bibinfo {author} {\bibfnamefont {J.}~\bibnamefont
  {Carrasquilla}}, \bibinfo {author} {\bibfnamefont {L.}~\bibnamefont
  {Taddia}}, \bibinfo {author} {\bibfnamefont {E.}~\bibnamefont {Ercolessi}}, \
  and\ \bibinfo {author} {\bibfnamefont {M.}~\bibnamefont {Rigol}},\ }\href
  {\doibase 10.1103/PhysRevB.91.165136} {\bibfield  {journal} {\bibinfo
  {journal} {Phys. Rev. B}\ }\textbf {\bibinfo {volume} {91}},\ \bibinfo
  {pages} {165136} (\bibinfo {year} {2015})}\BibitemShut {NoStop}%
\bibitem [{\citenamefont {Di~Liberto}\ \emph {et~al.}(2016)\citenamefont
  {Di~Liberto}, \citenamefont {Recati}, \citenamefont {Carusotto},\ and\
  \citenamefont {Menotti}}]{DiLiberto2016}%
  \BibitemOpen
  \bibfield  {author} {\bibinfo {author} {\bibfnamefont {M.}~\bibnamefont
  {Di~Liberto}}, \bibinfo {author} {\bibfnamefont {A.}~\bibnamefont {Recati}},
  \bibinfo {author} {\bibfnamefont {I.}~\bibnamefont {Carusotto}}, \ and\
  \bibinfo {author} {\bibfnamefont {C.}~\bibnamefont {Menotti}},\ }\href
  {\doibase 10.1103/PhysRevA.94.062704} {\bibfield  {journal} {\bibinfo
  {journal} {Phys. Rev. A}\ }\textbf {\bibinfo {volume} {94}},\ \bibinfo
  {pages} {062704} (\bibinfo {year} {2016})}\BibitemShut {NoStop}%
\bibitem [{\citenamefont {Tsukano}\ and\ \citenamefont
  {Nomura}(1998{\natexlab{a}})}]{tsukano1998berezinskii}%
  \BibitemOpen
  \bibfield  {author} {\bibinfo {author} {\bibfnamefont {M.}~\bibnamefont
  {Tsukano}}\ and\ \bibinfo {author} {\bibfnamefont {K.}~\bibnamefont
  {Nomura}},\ }\href@noop {} {\bibfield  {journal} {\bibinfo  {journal}
  {Journal of the Physical Society of Japan}\ }\textbf {\bibinfo {volume}
  {67}},\ \bibinfo {pages} {302} (\bibinfo {year}
  {1998}{\natexlab{a}})}\BibitemShut {NoStop}%
\bibitem [{\citenamefont {Tsukano}\ and\ \citenamefont
  {Nomura}(1998{\natexlab{b}})}]{tsukano1998spin}%
  \BibitemOpen
  \bibfield  {author} {\bibinfo {author} {\bibfnamefont {M.}~\bibnamefont
  {Tsukano}}\ and\ \bibinfo {author} {\bibfnamefont {K.}~\bibnamefont
  {Nomura}},\ }\href@noop {} {\bibfield  {journal} {\bibinfo  {journal}
  {Physical Review B}\ }\textbf {\bibinfo {volume} {57}},\ \bibinfo {pages}
  {R8087} (\bibinfo {year} {1998}{\natexlab{b}})}\BibitemShut {NoStop}%
\bibitem [{\citenamefont {Hayward}\ \emph {et~al.}(2018)\citenamefont
  {Hayward}, \citenamefont {Schweizer}, \citenamefont {Lohse}, \citenamefont
  {Aidelsburger},\ and\ \citenamefont {Heidrich-Meisner}}]{Hayward2018}%
  \BibitemOpen
  \bibfield  {author} {\bibinfo {author} {\bibfnamefont {A.}~\bibnamefont
  {Hayward}}, \bibinfo {author} {\bibfnamefont {C.}~\bibnamefont {Schweizer}},
  \bibinfo {author} {\bibfnamefont {M.}~\bibnamefont {Lohse}}, \bibinfo
  {author} {\bibfnamefont {M.}~\bibnamefont {Aidelsburger}}, \ and\ \bibinfo
  {author} {\bibfnamefont {F.}~\bibnamefont {Heidrich-Meisner}},\ }\href
  {\doibase 10.1103/PhysRevB.98.245148} {\bibfield  {journal} {\bibinfo
  {journal} {Phys. Rev. B}\ }\textbf {\bibinfo {volume} {98}},\ \bibinfo
  {pages} {245148} (\bibinfo {year} {2018})}\BibitemShut {NoStop}%
\end{thebibliography}%

\end{document}


\title{Supplementary material for "Topological charge pumping of bound bosonic pairs"}
\author{Sebastian Greschner}
\affiliation{Department of Quantum Matter Physics, University of Geneva, 1211 Geneva, Switzerland}
\author{Suman Mondal}
\author{Tapan Mishra}
\affiliation{Department of Physics, Indian Institute of Technology, Guwahati-781039, India}

\date{\today}

\begin{abstract}
In this Supplementary material we present a detailed discussion on the topological phase transitions of the attractive bosons at unit filling in the context of the SSH model which is 
a limiting case of the generalized RM model. First we discuss the qualitative 
picture of the system using the bosonization technique and then provide a detailed DMRG calculation to obtain various physical quantities to obtain the complete phase diagram including the calculation of 
the topological winding numbers. Further we present an elaborate discussion on the charge pumping and in the end we provide details on the experimental feasibility of these findings.  
\end{abstract}

\maketitle

\section{Ground state phase diagram at unit filling}
Detailed ground state bulk properties of three-body constrained bosonic SSH model in a grand canonical ensemble are discussed in Ref.~\cite{Mondal2019}. Here, we will summarize the phases and 
phase transitions of the model focusing on the phase diagram at unit-filling $n=N/L=1$ particles per lattice site. Before going to the detailed numerical results we will present the inferences that 
can be obtained from analytical arguments. The properties of (constrained) bosons at unit filling can be best understood from field theoretical arguments. 
In particular the system of 3-body constrained bosons can be understood as a spin-$1$ system, which itself can be described well by the triplet state of 
spin-$1/2$ states~\cite{Schulz1986}. Spin-1 models with staggered/dimerized exchange term have been studied before e.g. 
in Ref.~\cite{kitazawa1996phase, kitazawa1997critical, ejima2018exotic}. Here we will follow the bosonization scheme discussed e.g. by 
Berg et al. in Ref.~\cite{Berg2008}, and approximate the bosons by $a_j^{(\dagger)} \approx (a_{j,0}^{(\dagger)} + a_{j,1}^{(\dagger)})/\sqrt{2}$, 
with two different bosonic species $a_{j,\alpha}$, $\alpha=0,1$. Within the standard bosonization dictionary~\cite{GiamarchiBook}, we introduce two 
sets of bosonic fields ($\theta_{0},\phi_{0}$) and ($\theta_{1},\phi_{1}$) corresponding to density and phase fluctuations of $a_{j,\alpha}$. 
With Ref.~\cite{Berg2008} we result in an effective decoupling of symmetric and anti-symmetric 
combinations $\theta_\pm = (\theta_0 \pm \theta_1)/2$ and $\phi_\pm = \phi_0 \pm \phi_1$, $\mathcal{H} \approx \mathcal{H}_+ + \mathcal{H}_-$ and

\begin{align}
\mathcal{H_+} \sim \frac{ v_{+}}{2}\left[ \frac{(\partial_x \phi_{+})^2}{ K_{+}}+ K_{+}(\partial_x \theta_{+})^2  \right]  + g_1 \cos(2\phi_+) + \cdots
\end{align}
and
\begin{align}
\mathcal{H_-} \sim \frac{ v_{-}}{2}\left[ \frac{(\partial_x \phi_{-})^2}{ K_{-}}+ K_{-}(\partial_x \theta_{-})^2  \right]  +  g_2 \cos(2\phi_-) +\nonumber\\
+ g_3 \cos(2\theta_-) + \cdots
\end{align}
up to higher order terms. In weak coupling $g_{1,2} \sim U$ and $g_3 \sim t$ and phenomenological 
Luttinger-liquid parameters $K_\pm$, and velocities $v_\pm$. $\mathcal{H_-}$ has the form of a double-sine-gordon 
Hamiltonian~\cite{Takayoshi2018}. For $K_->1/2$, the $\cos(2\theta_-)$ term is relevant and will open up a gap in the antisymmetric sector. 
For strong attractive interactions $U<0$, the system may enter a phase of paired bosons where the $\phi_-$ field becomes gapped, 
which corresponds to the SF-PSF phase transition if the symmetric sector remains gapless. The $g_1$ term may induce a transition to a 
gapped phase in the symmetric sector, which corresponds to the opening of the Mott-gap and a MI-SF BKT like transition. In this framework 
the SSH-model staggered tunneling term $(-)^j \delta t a_j^\dagger a_{j+1}$ is given by
\begin{align}
\sim \delta t \cos(\phi_-) \sin(\phi_+) \,.
\label{eq:bos_stun}
\end{align}
In order to get some intuition of the physics of this term we may assume a mean-field like decoupling as $\sim \lambda_1 \cos(\phi_-) + \lambda_2 \sin(\phi_+)$, 
with $\lambda_1 = \la \sin(\phi_+) \ra$ and $\lambda_2 = \la \cos(\phi_-) \ra$. In the paired phase, we would expect to find $\lambda_2\neq 0$ and, 
hence, as $\sin(\phi_+)$ is a relevant term, it will here open up a gap also in the symmetric sector, introducing a phase transition to the PBO phase.
For strong $\delta t$ the $\cos(\phi_-)$ may directly open up a gap the antisymmetric sector also at vanishing $U$. Interestingly, both $\sin(\phi_+)$ and $\cos(2\phi_+)$ terms 
stabilizing PBO and MI phases are compatible with the pinning of $\phi_+\sim \pi/4$, and we may not expect any phase transition between the large-$U$ MI state and 
the PBO state, but a smooth crossover between the two regimes.

\begin{figure}[tb]
  \centering
  \includegraphics[width=0.99\linewidth]{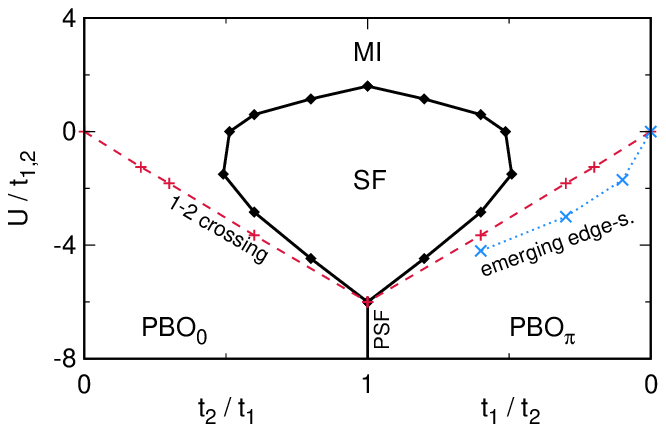}
    \caption{Phase diagram of the bosonic 3-body constraint SSH model at unit filling $n=1$ as function of $t_2/t_1$ and $U/t_1$ (on the left) and  $t_1/t_2$ and $U/t_2$ (right part). The straight 
    lines mark BKT phase transitions between gapped and gapless phases. The dotted line depicts the crossing
    between single and two particle excitations and defines the crossover between MI and PBO phases. 
    The dash-dotted line depicts the emergence of polarized edge states due to the nontrivial effective topology of the model in the attractive regime (see text).}
    \label{fig:n1_pd}
\end{figure}
\begin{figure}[tb]
  \centering
  \includegraphics[width=0.99\linewidth]{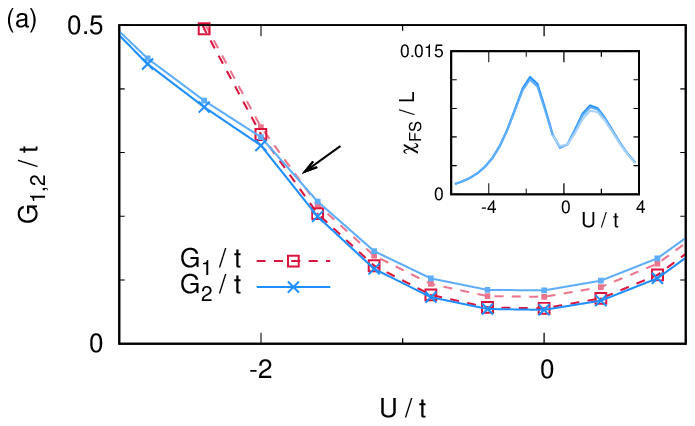}
  \includegraphics[width=0.99\linewidth]{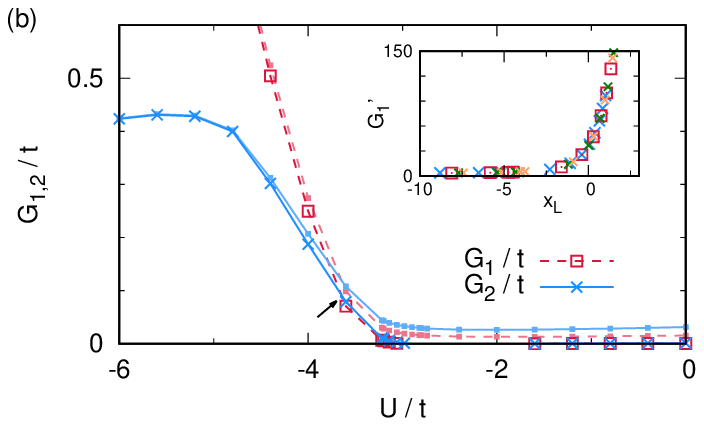}
    \caption{Single and two particle energy gaps $G_1$(red dashed lines) and $G_2$ (blue solid lines) are shown as function of $U/t$. We show the data for system size of $L=160$ (light filled squares)
    together with the extrapolation to the thermodynamic limit (open square for $G_1$ and cross for $G_2$) 
    using $L=40$, $80$, $120$ and $160$ sites). DMRG data for (a) $t_2/t_1=0.3$ and (b)  $t_2/t_1=0.6$. 
    The arrow marks the crossing between $G_1$ and $G_2$. The Inset in (a) depicts the fidelity susceptibility $\chi_{FS}/L$ for $t_2/t_1=0.3$ as function of $U$. 
    The Inset in (b) shows the BKT-compatible scaling of the single particle energy gap with a critical value of $U_c=-2.9 t_1$ (see text).}
    \label{fig:n1_gap}
\end{figure}

This feature is seen in the numerical calculation as well using the DMRG method. The complete ground state phase diagram of the bosonic constrained SSH model is shown 
in Fig.~\ref{fig:n1_pd} as a function 
of the hopping ratios $t_2/t_1$ as well as $t_1/t_2$ and the interaction strength $U/t_1$ and $U/t_2$ respectively. 
Ground state properties of this system is analyzed using the density matrix renormalization group~(DMRG) method. We consider typically system 
sizes up to $160$ sites and retaining up to $800$ density matrix eigenstates. 
The phase diagram shows various quantum phases such as the gapped MI phase on the repulsive
$U$ regime and the PBO$_0$ and PBO$_\pi$ phases in the attractive regime, which will be discussed in more detail in the following. 
Apart from this there also exist the gapless superfluid~(SF) and pair-superfluid~(PSF) phases. 
The black diamonds depict the phase transition to the SF phase and there exists smooth crossovers from the MI to the PBO phase denoted by the red plus-dashed lines. The blue cross-dotted line represents 
the emergence of edge states.

\begin{figure}[t]
  \centering
  \includegraphics[width=0.99\linewidth]{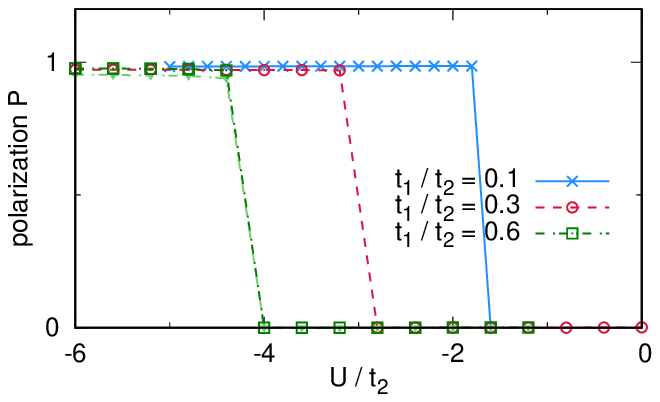}
    \caption{ Polarization $P$ for $t_1/t_2=0.1$, $0.3$ and $0.6$ as function of the interaction strength for $L=160$ and $L=80$ sites (light colored lines).
     }
    \label{fig:n1_edge}
\end{figure}

In Fig.~\ref{fig:n1_pd} the gap to gapless phase transitions and crossovers are determined by using the finite size scaling of the momentum distribution function;
\begin{equation}
 N(k)=\frac{1}{L}\sum_{i,j}e^{ik|i-j|}\la a_i^\dagger a_j\ra
\end{equation}
The BKT type MI-SF transition points are obtained by means of the finite size scaling of the momentum distribution function~\cite{Dhar2012,Greschner2013,Mondal2019} . 
We further compute the excitation gap of the single particle $G_{1}$ and two-particle sector $G_{2}$ with 
\begin{equation}
G_{m}(L) =(E_0(N+m,L) - 2 E_0(N,L) + E_0(N-m,L))/m,
\end{equation}
where $E_0(N,L)$ is the ground-state energy of a system of $N$ particles and $L$ sites and $m\in[1,~2]$.
In order to verify the BKT character of the phase transitions we analyze the finite size 
scaling of the energy gaps in the negative $U$ regime using the approach used in Ref.~\cite{Mishra12, Dalmonte14}. In the inset of Fig.~\ref{fig:n1_gap} the rescaled single 
particle energy gap $G'_1 = G_1 L (1+1/(\log(L)+C))$ for $t_2=0.6$ exhibits the expected scaling close to the BKT transition
point($U_c=-2.9$) leading to a collapse of data points plotted against $x_L=\log{L} - \log{B/\sqrt{|U-U_c|}}$, with fitting parameters 
$C$, $B$ and $U_c$ ~\cite{Mishra12, Dalmonte14}. 
For a strong staggered hopping $\delta t$ as shown in Fig.\ref{fig:n1_gap}~(a) we do not observe any closing of the excitation gap $G_{1,2}$ upon changing $U$ from the PBO to the MI phase. 
Both phases remain adiabatically connected. As discussed already in Ref.~\cite{Mondal2019} the fidelity susceptibility and parity order stays finite during this 
crossover between different regions. In the inset of  Fig.\ref{fig:n1_gap}~(a) we depict the 
fidelity susceptibility defined by 
\begin{equation}
 \chi_{FS}(U) = \lim_{U-U' \to 0} \frac{-2 \ln |\langle \psi_0(U) |\psi_0(U') \rangle| }{(U-U')^2},
\end{equation}
where $|\psi_0\rangle$ is the ground-state wave-function. $\chi_{FS}/L$ exhibits local maximum which stays finite as function 
of the system size. The presence of a second local maximum may be attributed to an intermediate bond-ordered~(BO) region~\cite{Mondal2019}, which we will not discuss here further.
Numerically, we may quantify the crossover points from the  MI-region to the PBO phase by an analysis of the excitation gap. 
As shown in Fig.~\ref{fig:n1_gap}~(a) and (b) we observe a crossing between the single particle gap $G_{1}$(red dashed lines) and two-particle gap $G_{2}$ (blue solid lines).
In the figure we show the data for $L=160$( along with the extrapolated data for $t_2=0.3$ (Fig.~\ref{fig:n1_gap}~(a)) and $0.6$ (Fig.~\ref{fig:n1_gap}~(b)). 
The extrapolation  is performed using the data for $L=40,~80,~120,~160$. The arrow defines the crossover between MI-like and dimerized PBO states. 
The crossover position to a PBO region can also be seen in the bond order structure factors~\cite{Mondal2019}.

\section{Topological properties}

\subsection{Edge states}
As discussed in the main text the PBO$_\pi$ phase exhibits non trivial edge states. 
In order to quantify the edge properties in more detail, we calculate the polarization 
\begin{align}
P = \frac{1}{L} \sum_{i=0}^{L} \langle \psi | (i-i_0) n_i | \psi \rangle
\end{align}
with $i_0=(L-1)/2$ and the ground state $|\psi\rangle$ of Model~\eqHRM or ~\eqSSH.
We plot this quantity for different values of $t_1/t_2$ with respect to $U$ as shown in Fig.~\ref{fig:n1_edge}. 
In order to facilitate the numerical simulations we add a small symmetry breaking potential to the boundary sites, 
corresponding to the evolution of the RM model at small value of $\delta \tau=0.001$. Comparisons to results with a 
smaller $\delta\tau=0.0001$ and different system sizes indicate that our choice does not influence the physics of the system. 
Interestingly, we observe a sudden sharp transition between a region of $U<U_e$ with polarized edges and a polarization close to $P=1$ and unpolarized systems with $P=0$ for larger values of $U>U_e$. 

In the phase diagram Fig.~\ref{fig:n1_pd} we depict the point of emergence of edge states $U_e$ as the blue cross-dotted line. This transition line differs 
from the crossing position of 2-particle and 1-particle excitations (red dashed line in Fig.~\ref{fig:n1_pd}). The reason for this can be attributed to the bosonic enhancement 
which is energetically favorable to first delocalize a single doublon on the MI phase before the polarized edge states are formed~\cite{DiLiberto2016}.

\begin{figure}[t]
  \centering
  \includegraphics[width=0.49\linewidth]{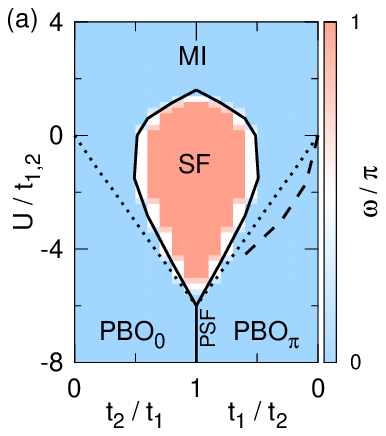}
  \includegraphics[width=0.49\linewidth]{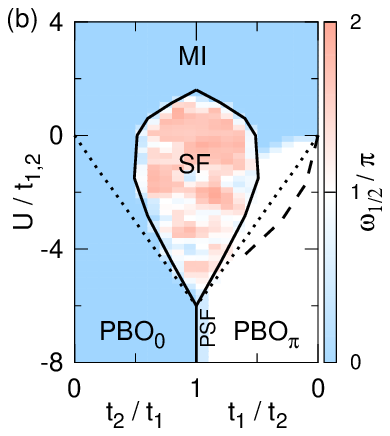}
    \caption{Estimates of the winding numbers (a) $\omega$ and (b) $\omega_{1/2}$ from small system 
    size simulation ($L=8$ sites, exact diagonalization) for the SSH-model as function of $U/t$ and $t_1/t_2$ as well as $t_2/t_1$. Black lines correspond to the lines of Fig.~\ref{fig:n1_pd}.}
    \label{fig:n1_wnr}
\end{figure}

\subsection{Winding number $\omega$}
In this section we compute the winding number using the standard definition in the many-body context from the ground state $|\psi\rangle$ of the 
effective model with twisted boundary conditions $a_{i} \to {\rm e}^{i\theta / L} a_i$,
\begin{align}
\omega = \int_0^{2\pi} d\theta \langle \psi(\theta) |\partial_\theta \psi(\theta) \rangle
\label{eq:winding1}
\end{align}
With this definition we find that  at unit-filling $\omega$ vanishes in all the gapped phases. 
The SF phase is accurately characterized by the winding number $\omega\neq 0$ corresponding to its superfluid density. 
However, we observe no distinction between the $t_1<t_2$ and the  $t_2<t_1$ region as shown in Fig.~\ref{fig:n1_wnr}~(a). This problem was countered by redefining $\omega$ as 
discussed in the main text and as a result we reproduce the desired values as shown in Fig.~\ref{fig:n1_wnr}~(b).

\subsection{Charge Pumping in the Rice-Mele model}
As a starting point we may also obtain insights into the physics of the RM model at unit-filling by means of the 
bosonization treatment discussed above. Analogously to Eq.~\eqref{eq:bos_stun} we may express the staggered potential as
\begin{align}
\sim \delta \Delta \cos(\phi_-) \cos(\phi_+)
\label{eq:bos_spot}
\end{align}

For a pure staggered potential we may, hence, expect a competition between the MI phase of 
dominant $\cos(2 \phi_+)$ in the large $U$ limit and the DW ordering $\cos(\phi_+)$ and observe phase 
transition between both phases. For the case of the related spin-$1$ system with a staggered field $(-1)^x S^z_x$ 
this has been studied in detail in Refs.~\cite{tsukano1998berezinskii,tsukano1998spin} and a Gaussian type transition 
between large-U (which corresponds to the MI phase in the bosonic system) phase and N\'eel ordered phase (corresponding to the DW phase) has been observed. 

The effect of both a staggered potential and staggered tunneling term as we will encounter during the pumping cycle, can be expressed as a single cosine-term
\begin{align}
\sim \cos(\phi_-) \cos(\phi_+ + \varphi) \,,
\label{eq:bos_RM}
\end{align}
with $\varphi \sim \arctan(\delta t / \delta \Delta) \sim \tau$. The value of the pinned field $\phi_+$ will thus follow smoothly the 
variations of $\tau$. In a finite system this will correspond to pumping of a charge. The competition with MI term should lead again 
to interesting crossover or transitions which we discuss in in the main text by a more detailed numerical treatment.

\begin{figure}[t]
  \centering
  \includegraphics[width=0.49\linewidth]{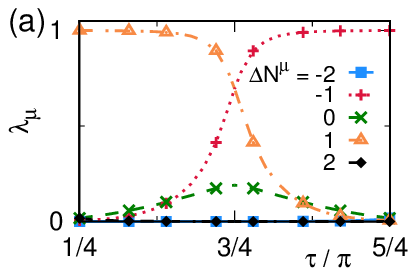}
  \includegraphics[width=0.49\linewidth]{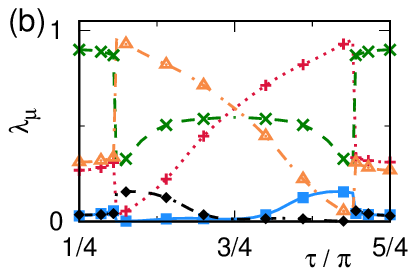}
  \includegraphics[width=0.49\linewidth]{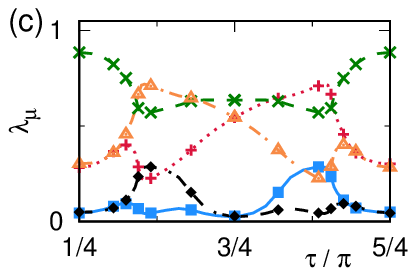}
  \includegraphics[width=0.49\linewidth]{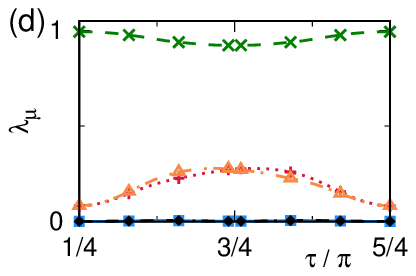}
    \caption{ Largest values of the entanglement spectrum $\lambda_\mu$ of the reduced density matrix in the center of a finite system of size $L=160$ 
    for the RM model shown in Eq.~\eqHRM of the main text as a function of the adiabatic parameter $\tau$ for (a) $U=-1$ and (b) $U=4$ (compare Ref.~\cite{Hayward2018}). 
    The eigenvalues $\lambda_\mu$ have been sorted by the particle number sector $N_\mu$ in the center of the chain and for convenience labeled by $\Delta N_\mu = N_\mu - L/2$.	
     }
    \label{fig:n1_svpump}
\end{figure}
As discussed in the main text the evolution of the entanglement spectrum $\lambda_\mu$ can provide signatures of charge pumping. In Fig.~\ref{fig:n1_svpump} we plot $\lambda_\mu$ for several 
values of $U$. With increasing $U$ we observe a smooth crossover to the MI regime where the $\Delta N_\mu=0$ state has the largest contribution and no charge is pumped. 

\begin{figure}[t]
  \centering
  \includegraphics[width=0.49\linewidth]{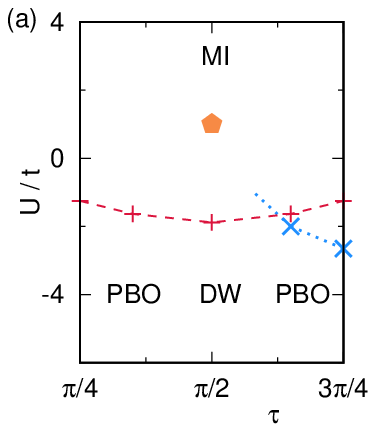}
  \includegraphics[width=0.49\linewidth]{pd_pmpRM_JA0.9_JB2.0.eps}
    \caption{Charge pumping in Model~\eqHRM  ~of the main text for (a) $t_1=0.2 t_2$, $\delta \Delta=\delta t =1 $ (b) $\delta t=0.9$, $\delta \Delta=2$. 
    The crosses mark the breakdown of pair-pumping as seen by the sharp kink in the polarization. 
    As discussed before, this only qualitatively coincides with the crossing position of two- and single 
    particle excitations (red plus symbols). Only for $\tau=1/4$ we observe a gapless phase transition point 
    between a density wave(DW) and a MI like region (orange hexagon). Also in the vicinity of this point the charge gap becomes strongly suppressed.}
    \label{fig:pd_pmpRM}
\end{figure}

As a topological feature charge pumping should be robust with alterations of the actual pumping protocol or
periodic path chosen through the parameter space of the RM model. In Fig.~\ref{fig:pd_pmpRM}~(a) we repeat 
the analysis shown in Fig.~2(b) of main text for a more elongated path with $\delta t = \delta \Delta = 1$. We again observe 
pumping of bosonic pairs for strong attractions of $-U \lesssim 2.6 t$. In Fig.~\ref{fig:pd_pmpRM} we show both the phase diagrams for comparison. 

\begin{figure}[t]
  \centering
  \includegraphics[width=0.99\linewidth]{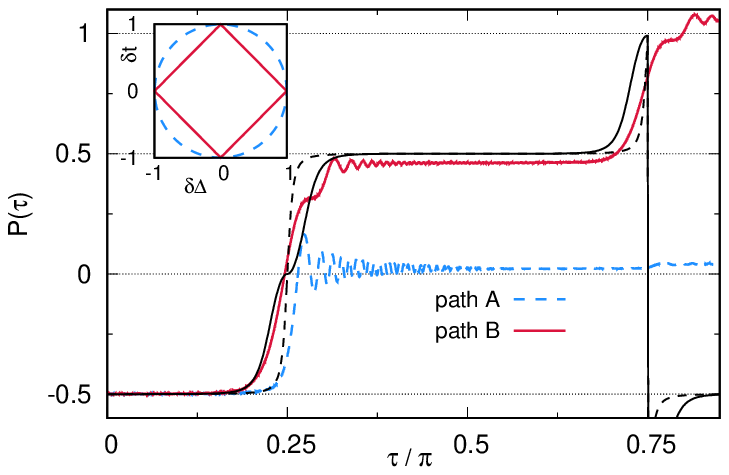}
    \caption{Pumping of repulsively bound pairs along different paths A and B (sketched in the inset as parametric plot in $\tau$). 
    The dashed lines depict the fully adiabatic evolution (ground state calculation of the effective model of attractively bound pairs, $L=120$ sites). 
    The solid lines show the simulation of an actual real time evolution with $t = 10 \tau$, which is sufficiently slow to stay close to the adiabatic 
    regime in path $B$ but too fast for path $A$ (exact diagonalization, $L=6$ sites, $N=6$ particles).}
    \label{fig:tevP_Uneg_sc}
\end{figure}
\section{Time evolution}

In Fig.~\ref{fig:tevP_Uneg_sc} we perform the time dependent simulation of the pumping 
process after a quench of the interaction $U\to U_q$ and compare to the pumping of attractively bound pairs. Generally we choose a 
relatively fast evolution $t = 10 \tau$. Interestingly, if we change the precise path through the phase space (path B as sketched in 
the inset of Fig.~\ref{fig:tevP_Uneg_sc}) the (partial) pumping cycle can be performed accurately as shown in Fig.~\ref{fig:tevP_Uneg_sc}. 
Instead of the elliptical path of Eq.~(1) of main text we choose a more rectangular contour (path B), connecting still the same intermediate 
points at $\tau\,{\rm mod }\, \pi/4$, our real time evolution is able to pump a repulsively bound pair, while on path A, the evolution is 
too fast and the system evolves to an unpolarized state.

\bibliography{references}